\RequirePackage{ifpdf}
\documentclass[hyper,12pt,a4paper]{JHEP3}
\pdfoutput=1
\usepackage{cite}
\usepackage{psfrag}
\usepackage{graphicx}
\usepackage{enumitem}
\usepackage{bm}
\usepackage{amsmath}
\usepackage{float}

\usepackage{lipsum}

\usepackage{amssymb}
\usepackage{fancyhdr}
\usepackage{url}
\usepackage{epstopdf}
\usepackage[utf8]{inputenc}
\usepackage{verbatim}
\newcommand{\Tr}{\text{Tr}}

\newcommand{\seff}[1]{S_{\text{eff}}({#1})}

\newcommand{\ben}{\begin{eqnarray}\displaystyle}
\newcommand{\een}{\end{eqnarray}}

\newcommand{\be}{\begin{equation}}
\newcommand{\ee}{\end{equation}}

\newcommand{\bc}{\begin{center}}
\newcommand{\ec}{\end{center}}

\newcommand{\eesp}{\end{split}}
\newcommand{\bsp}{\begin{split}}

\newcommand{\Rmnum}[1]{\expandafter\@slowromancap\romannumeral #1@}
\renewcommand{\a}{\alpha}	%%% Redefinition
\renewcommand{\b}{\beta}		%%% Redefinition
\newcommand{\dow}{\partial}
\renewcommand{\l}{\lambda}	%%% Redefinition
\renewcommand{\o}{\omega}	%%% Redefinition
\newcommand{\q}{\theta}	
	
\renewcommand{\r}{\rho}		%%% Redefinition
\newcommand{\bo}{\beta_{1}}

\newcommand{\cD}{\mathcal{D}}

\newcommand{\cM}{\mathcal{M}}

\newcommand{\cO}{\mathcal{O}}

\newcommand{\cZ}{\mathcal{Z}}

\newcommand{\expb}[1]{\exp\left[ #1 \right]}
\newcommand{\zcs}{\cZ_{\text{CS}}}

\newcommand{\lnb}[1]{\ln\left[ #1 \right]}

\newcommand{\ra}{\rightarrow}

\newcommand{\lB}{\left [}
\newcommand{\rB}{\right ]}
\newcommand{\lb}{\left (}
\newcommand{\rb}{\right )}

\renewcommand{\>}{\right\rangle}		%%% Redefinition

\newcommand{\with}{\text{with}}

\newcommand{\bensp}{\begin{eqnarray}\begin{split}}
\newcommand{\eensp}{\end{eqnarray}\end{split}}

\newcommand{\bnm}{\begin{matrix}}
\newcommand{\enm}{\end{matrix}}
\def\Xint#1{\mathchoice
{\XXint\displaystyle\textstyle{#1}}%
{\XXint\textstyle\scriptstyle{#1}}%
{\XXint\scriptstyle\scriptscriptstyle{#1}}%
{\XXint\scriptscriptstyle\scriptscriptstyle{#1}}%
\!\int}
\def\XXint#1#2#3{{\setbox0=\hbox{$#1{#2#3}{\int}$ }
\vcenter{\hbox{$#2#3$ }}\kern-.6\wd0}}

\newcommand{\half}[1]{{#1\over 2}}
\newcommand{\dgr}{\dagger}

\newcommand{\gt}{\theta}
\newcommand{\gl}{\lambda}
\newcommand{\ga}{\alpha}
\newcommand{\gb}{\beta}

\newcommand{\gep}{\epsilon}

\newcommand{\stso}{$S^2\times S^1$}

\usepackage{float}
\usepackage[font=small,labelfont=bf]{caption}
\usepackage{mathrsfs}
\usepackage[parfill]{parskip}
\usepackage{subcaption}
\graphicspath{{chobi/}}

\numberwithin{equation}{section}
\usepackage{tabularx}

\usepackage{youngtab} % for Young tableaux
\usepackage{young} % for Young tableaux
\title{From Phase Space to Integrable Representations and Level-Rank Duality} 
\author{Arghya Chattopadhyay\footnote{arghya@iiserb.ac.in} $^{1}$,
	Parikshit Dutta\footnote{parikshitdutta@yahoo.co.in} $^{2}$\ and Suvankar
	Dutta\footnote{suvankar@iiserb.ac.in} $^{1}$\\
	$^{1}$Department of Physics \\
	Indian Institute of Science Education and Research Bhopal \\
	Bhopal 462 066, 
	India\\
	
	$^{2}$Asutosh College\\
	92, Shyamaprasad Mukherjee Road\\
	Kolkata 700 026, India\\
	}
\abstract{We explicitly find representations for different large $N$ phases of Chern-Simons matter theory on $S^2\times S^1$. These representations are characterised by Young diagrams. We show that no-gap and lower-gap phase of Chern-Simons-matter theory correspond to integrable representations of $SU(N)_k$ affine Lie algebra, where as upper-cap phase corresponds to integrable representations of $SU(k-N)_k$ affine Lie algebra. We use phase space description of \cite{duttagopakumar} to obtain these representations and argue how putting a cap on eigenvalue distribution forces corresponding representations to be integrable. We also prove that the Young diagrams corresponding to lower-gap and upper-cap representations are related to each other by transposition under level-rank duality. Finally we draw phase space droplets for these phases and show how information about eigenvalue and Young diagram descriptions can be captured in topologies of these droplets in a unified way.}
\begin{document}

\section{Introduction and summary}\label{sec:intro}

Droplet description of $SU(N)$ gauge theories \cite{duttagopakumar} allows one to find dominant large $N$ representations in terms of Young diagram \cite{Chattopadhyay}. Information about different phases of gauge theory is encoded in the geometry or shape of droplets. From this geometry it is possible to construct corresponding Young diagrams and hence different large $N$ states of the theory in momentum space. A phase space description for large $N$ gauge theories is apparent when one writes gauge theory partition function in terms of unitary matrix models. Solution of these matrix models renders a distribution of eigenvalues (EVs) of holonomy matrix. EV distribution captures information of different phases of the system and transition between them. Eigenvalues of these matrices behave like positions of free fermions \cite{bipz}. On the other hand, writing the same partition function as a sum over all possible representations of unitary group, one can study large $N$ behaviour of this theory in terms of the most dominant representation which is characterised by a function $u(h)$ \cite{duttagopakumar,douglas-kazakov,Kazakov:1995ae}. $u(h)$ measures how boxes $h$ are distributed in a Young diagram. Solving the model at large $N$, \cite{duttagopakumar} found that information of different phases can be captured in terms of topologies (shapes) of different droplets in two dimensions. This two dimensional plane is spanned by $\q$ and $h$. In \cite{Chattopadhyay} it was argued that number of boxes play the role of momenta of underlying free fermi theory. Therefore, droplets in $(h,\q)$ plane are similar to those of Thomas-Fermi model where Wigner distribution is assumed to take constant value inside some region in phase space and zero elsewhere. 

In this paper we use phase space description to find out representations corresponding to different large $N$ phases of Chern-Simons-Matter (CSM) theory on $S^2\times S^1$ with gauge group $SU(N)$ and level $k_{YM}$ and study properties of these representations and relations between them in the context of {\it level-rank} duality, enjoyed by this theory.

Duality plays a vital role in physics. Two seemingly different theories sometimes are related to each other by duality transformations. For example, the most popular example of duality in recent years is the AdS/CFT duality, which relates a super-conformal quantum field theory (without gravity) to a superstring theory in one higher dimensions. However, our focus will be on supersymmetric or non-supersymmetric gauge theories. In general, dual theories may have different gauge groups or matter field representations or even number of degrees of freedom. One such example of this duality is the Seiberg duality \cite{seiberg1995}, which relates $\mathcal{N}=1$ $d=4$ supersymmetric gauge theories. Much like Seiberg duality, Giveon and Kutasov (GK) realised a duality in $\mathcal{N}=2$ $d=3$ Chern-Simons matter theories \cite{giveonkutasov}. The GK duality relates $U(N_c)_k$ CS theory with $N_f$ fundamental flavours with $U(|k|+N_f-N_c)_{-k}$ CS theory with $N_f$ fundamental flavours. The subscripts $k$ here denotes renormalised level of the theory defined as  $k=k_{YM}+N$. Where $k_{YM}$ is the level of the CS theory regulated by an infinitesimal Yang-Mills term in the action. For $N_f=0$ this duality can be shown to be reminiscent of "level-rank" duality in CS theory \cite{kapustintest}. The level-rank duality (exchange of $N\leftrightarrow k_{YM}$) was first observed by \cite{Naculich:1990} in the context of Wess-Zumino-Witten (WZW) model\footnote{See also \cite{Mlawer:1990uv,Naculich:1990hg}}. They showed the duality between primary fields of $SU(N)_{k_{YM}}$ and $SU(k_{YM})_N$ WZW model, which can also be extended to duality between correlation functions of this fields. For example if the primary fields of $SU(N)_{k_{YM}}$ WZW model transform in some integrable representation\footnote{More on this in can be found in \ref{app:yng}.} $Y$ of $SU(N)_{k_{YM}}$, characterised by Young diagram with at most ${k_{YM}}$ number of boxes in the first row, then in the dual theory, primary fields will transform under integrable representation $\tilde{Y}$ of $SU(k_{YM})_N$, where $Y$ and $\tilde{Y}$ are related by ``transposition", i.e, Young diagrams for these representations are related by interchange of rows and columns. Similarly for Chern-Simons theories the objects that enjoy this duality are Wilson loops. Representations of Wilson loops in both the theories are again related by `` transposition ".

We consider level $k$ (renormalised level) $U(N)$ CS theories on $S^2\times S^1$ coupled with fundamental fields in the 't Hooft limit\footnote{ ${U(N)=[SU(N)\times U(1)]/\mathbb{Z}_N}$.}. Interestingly it turns out that $U(1)^N$ flux sectors in this theory discretizes the eigenvalues of holonomy matrix. We shall discuss this in details in the next section. The discretization imposes a strict upper bound or cap of $1/ 2\pi\gl$ on eigenvalue density. Here $\gl$ is 't Hooft coupling : $\l=N/k$. An upper cap on eigenvalue distribution generates a rich phase structure compared to ordinary YM theories on compact spaces \cite{gross-witten, wadia, zalewski, AMMKR, Basu:2005pj, AlvarezGaume:2006jg, Mandal:1989ry, Yamada:2006rx, AlvarezGaume:2005fv, Harmark:2006di, Friedan:1980tu, Sundborg:1999ue}. Saturation of eigenvalue density was first suggested in \cite{giombi}. This particular kind of matrix model is dubbed as capped matrix model. Apart from usual ``no-gap" and ``one-gap" (or "lower gap") phases similar to YM theories, one has two new phases. The phase, where eigenvalues are distributed over unit circle (much like no-gap phase) but eigenvalue density attains the saturation limit on some finite arc. This phase is called "upper-cap" phase. The second new phase arises when eigenvalue distribution for a lower-gap phase saturates the limiting value over a finite arc on unit circle. We use the name "upper cap with lower gap" for this phase. \cite{shirazs2s1} studied this phase structure by considering {\it Gross-Witten-Wadia} (GWW) \cite{gross-witten, wadia} type potential and argued that actual CS matter theory will also exhibit same phase structure. \cite{shirazs2s1} also determined potentials for the capped matrix model corresponding to several CS matter theories alongside their saddle point equations. Later, \cite{takimi2013} extended this work and found exact solutions for different saddle point equations. Their results are also consistent with predictions of \cite{shirazs2s1}. Though thermal partition functions are not topological quantities, interestingly level-rank duality does establish equality between thermal partition functions of dual theories in high temperature limit. \cite{takimi2013} proved that interchanging level and rank actually maps lower gap phase of one theory to upper cap phase of its conjectured dual theory at high temperature. Some recent works on CSM theory can be found in \cite{Codesido:2014oua, Marino:2012az, Jain:2012qi, Dandekar:2014era, Jain:2013gza, Giombi:2011kc, Minwalla:2011ma, Suyama:2011yz}.

Work of \cite{Naculich:1990} motivates us to ask the question whether constraint on eigenvalue density imposes any conditions or restrictions on large $N$ representations where the fields transform. Prime goal of this paper is to address this question. We consider CS theory on $S^2\times S^1$ in presence of GWW potential. Salient observations in this paper are following.
\begin{itemize}
	\item Constraints on eigenvalue distribution impose restrictions on number of boxes in the first row of the most dominant representation in large $N$ limit.
	\item We find no-gap and lower gap phases correspond to integrable representations of $SU(N)_k$ theory where as, upper-cap phase corresponds to integrable representations of $SU(k-N)_k$.
	\item Eigenvalue densities for lower-gap phase and upper-cap phase are related to each other by level-rank duality \cite{shirazs2s1}. We observe that dominant large $N$ Young diagrams corresponding to these two phases are related to each other by transposition which is consistent with observation of \cite{Naculich:1990}.
	\item We find droplets in $(h,\q)$ plane corresponding to these phases and observe topological difference between the droplets corresponding to different large $N$ phases. 
	\item Information about eigenvalue and Young diagram distribution can be captured in geometries of these droplets in a unified way.
\end{itemize}

We expand the partition function for GWW matrix model in momentum (or Young diagram) basis \cite{duttagopakumar,riemannzero,Chattopadhyay,douglas-kazakov} with the additional constraint coming from CS theory. In large $N$ limit one can perform a saddle point analysis and find representations (i.e. profile of the Young diagrams) which dominate the partition function. \cite{duttagopakumar, riemannzero,Chattopadhyay} noticed a surprising identification between eigenvalue distribution and Young diagram distribution for different phases of a generic unitary matrix model. This identification allows one to provide a phase space description at large $N$. Different phases are characterised by topologies of these phase space droplets. Both the eigenvalue and Young diagram distributions can be obtained from the shape of these droplets by suitably integrating out appropriate degrees of freedom. We use this technique to find dominant representations for CS GWW matter theory.

%%%%%%%%%%%%%%%%%%%%%%%%%%%%%%%%%%%%%%%%%%%%%%%%%%%%%%%%%%%%%
%%%%%%%%%%%%%%%%%%%%%%%%%%%%%%%%%%%%%%%%%%%%%%%%%%%%%%%%%%%%%%%

In \cite{Wittenjones}, Witten showed that topological invariants of knots and links (Ex. Jones and HOMFLY ploynomial) can be rewritten as correlation functions of Wilson loop operators in Chern-Simons theory. This observation initiated a thorough investigation of Chern-Simons theory as a topological field theory. By working out the exact solution of Chern-Simons theory Witten further showed a connection of these knots and topological invariants of three manifold  with WZW model. In particular, if one quantises the Chern-Simons theory on a three manifold\footnote{$\Sigma_g$ is genus $g$ 2-manifold.} $\Sigma_g\times S^1\,\,$  then the Hilbert space $\mathcal{H}(\Sigma_g)$, associated with a two manifold $\Sigma_g$ , can be described as space of conformal blocks of WZW model on $\Sigma_g$ with some gauge group $G$ and level $k$. In fact, the level-rank duality of $U(N)$ or $SU(N)$ Chern-Simons theory actually follows from the level rank duality of WZW model with gauge group $U(N)$ \cite{Naculich:2007nc}. The affine lie algebra of the $U(N)$ WZW model is $\widehat{su}(N)_k\times \widehat{u}(1)_{k'}/\mathbb{Z}_N$. Different physical observables of CS-theory on $\Sigma_g\times S^1$ in a given representations of the gauge group $U(N)$ (or SU(N)) can be written in terms of the observables of WZW model in representations of affine gauge group of WZW. Since, CS theory is topological, its observables are topological invariants of the manifold $\mathcal{M}$. The partition function depends only on $\mathcal{M}$, the gauge group $G$, the Chern-Simons coupling $k$, and the choice of framing \cite{Wittenjones}. Topological invariants of knots or links can be written in terms of the gauge invariant Wilson loops, defined as the expectation of the path-ordered integrals $\text{Tr}_R P \exp (\oint_\mathcal{K} A)$ around a closed path or knot $\mathcal{K}$ in $\mathcal{M}$, where the trace is taken in the irreducible representation $R$ of $G$. A link $\mathcal{L}$ with components $\mathcal{K}_\alpha,\,\,\alpha=1,\cdots,L$ is defined as
	\begin{equation}
	W_{R_1\cdots R_L}=<W_{R_1}^{\mathcal{K}_1}\cdots W_{R_L}^{\mathcal{K}_L}>
	\end{equation}  
These observables can be expressed in terms of the observables of WZW in different representations of $\hat{g}_K$ \cite{Wittenjones, Naculich:2007nc}
\begin{equation}
W_{R_1\cdots R_n}[\Sigma_g\times S^1,G,K]=\sum_R S_{0R}^{2-n-g}\prod_{i=1}^n S_{RR_i},
\end{equation}
where $S_{RR_i}$\footnote{{The characters of the highest weight representations transforms into one another under the modular transformation\cite{yellowbook} $\tau\leftarrow-1/\tau$}: $\chi_R(-1/\tau)=\sum_{R'}S_{RR'}\chi_{R'}(\tau)$.} is	the modular transformation matrix of the $\hat{g}_K$ WZW model, The sum is over integrable representations of $\hat{g}_K$, and $0$ denotes the identity representation. In our case we are dealing with the Chern-Simons theory on $S^2\times S^1$, the Wilson lines can be expressed as 
 \begin{equation}
 W_{R}[S^2\times S^1,U(N),k]=\sum_R S_{0R}S_{RR_1}.
 \end{equation}
 Therefore in the large $N$ limit the representations of $CS$ theory should have some relation with the integral representation of the corresponding affine algebra. In this paper, to our surprise we observe that in the large $N$ limit the dominant representations of $SU(N)$ turns out to be integrable representations of $\widehat{su}(N)_k$ of WZW model. We are yet to understand the deep insight behind this observation.   
 
 %%%%%%%%%%%%%%%%%%%%%%%%%%%%%%%%%%%%%%%%%%%%%%%%%%%%%%%%%%%%%%%%%%%%%%%%%%%
 %%%%%%%%%%%%%%%%%%%%%%%%%%%%%%%%%%%%%%%%%%%%%%%%%%%%%%%%%%%%%%%%%%%%%%%%%%%

Organisation of our paper is following.
\begin{enumerate}
	\item In section \ref{sec:review} we present a short review of eigenvalue analysis for CS theory in presence of GWW potential.
	\item We discuss phase space description for unitary matrix model in presence of GWW potential in section \ref{sec:youngreview}. We explain how one obtains the relation between eigenvalue description and Young diagram description in this section.
	\item The main observation of this paper has been presented in section \ref{sec:constraint_rep}.
	\item We end this paper with a vivid discussion and outlook in section \ref{sec:discussion}.
	\item Appendix \ref{app:yng} discusses representation of affine Lie algebras.
\end{enumerate}

\section{Eigenvalue analysis}\label{sec:review}
Chern-Simons theory is an example of a ``Schwartz-type" topological field theory, which is characterised by its metric independence. The partition function of a level $k$ $U(N)$ Chern-Simons theory on any three-dimensional manifold $\mathcal{M}$ is given by
\begin{eqnarray}
\zcs=\int_\cM [\cD A] \expb{{ik\over 4\pi}\Tr \int (A\wedge dA+{2\over 3}A\wedge A\wedge A)}.
\end{eqnarray} 
The topological invariants of this theory are the correlation functions of the Wilson loops, defined as the trace of the holonomy of $A$ around some closed curve $\gamma$ as the following
\begin{equation}
W_\gamma(A)=\Tr P\,\,exp\int_\gamma A
\end{equation}
where $P$ stands for path ordering. 

We consider CS theory on $S^2\times S^1$ interacting with matter in fundamental representations. Partition function of this theory is given by
\begin{eqnarray}
\cZ=\int [\cD A] [\cD \mu]e^{i{k\over 4\pi}\Tr\int(AdA+{2\over 3}A^3)-S_{matter}},
\end{eqnarray}
where $D\mu$ is the matter field measure. Now the job is to integrate out the matter fields and obtain some effective action which only depends on gauge fields. Following \cite{shirazs2s1}, the partition function is given by
\begin{eqnarray}
\cZ=\int [\cD A]e^{i{k\over 4\pi}\Tr\int(AdA+{2\over 3}A^3)-S_{eff}(U)},
\end{eqnarray}
where $U(x)$ is the two dimensional holonomy field around the thermal circle $S^1$. The eigenvalues of this holonomy matrix are defined as $e^{i\ga_m(x)}$ where $m$ runs from $1$ to $N$. The effective action $S_{eff}(U)$ can be obtained by summing over all the vacuum Feynman diagrams which includes at least one matter field. The locality of $S_{eff}$ in high temperature can be inferred from the observation that all matter fields acquire thermal masses and to obtain $S_{eff}$ we have to integrate out those diagrams which always include atleast one matter propagator. Now if we introduce the parametrisation
\begin{eqnarray}\label{scaling}
V_2T^2=N\gb_1
\end{eqnarray}
where $V_2$ is the proper volume of the spatial manifold $S^2$, then following \cite{giombi,shirazs2s1} one may expand $S_{eff}(U)$ as a series of local operators as
\begin{eqnarray}
S_{eff}(U)=\int d^2x(T^2\sqrt{g} v(U)+v_1(U)\Tr D_i UD^iU+...).
\end{eqnarray}
Equation (\ref{scaling}) is such that it converts this expansion into a series in ${1\over N}$. Then at large $N$ the thermal partition function for the Chern-Simons matter theories is given by
\begin{eqnarray}\label{eq:fullpartition}
\cZ=\int [\cD A] \expb{{ik\over 4\pi}\Tr \int (A\wedge dA+{2\over 3}A\wedge A\wedge A)-T^2\int d^2x\sqrt{g}v(U(x))}.
\end{eqnarray} 
This partition function is same as a pure CS theory with an additional term. This extra term represents the entire effect of the matter loops in CS matter theories at temperature $\sqrt{N}$ and at leading order in $N$. Because of the topological invariance of pure CS theory, expectation values in the pure CS theory is always independent of $x$. Therefore, the partition function can be rewritten as
\begin{eqnarray}
\cZ=\<e^{-T^2V_2v(U)}\>_{N,k},
\end{eqnarray}  
where the expectation is calculated in the pure CS theory with rank $N$ and level $k$. Generalizing the method of \cite{blauthompson}, \cite{shirazs2s1} evaluated this path integral and showed that it can be written as a summation over the ``discrete" eigenvalues of the holonomy matrix $U$ as
\begin{eqnarray}
\zcs=\prod_{m=1}^N\sum_{n_m=-\infty}^\infty\left[\prod_{l\neq m}2 \sin\left( {\gt_l(\vec{n})-\gt_m(\vec{n})\over 2}\right)e^{-V(U)} \right].
\end{eqnarray}
with
\begin{equation}
V(U)=T^2V_2v(U),\quad \gt_m(\vec{n})={2\pi n_m\over k}\qquad n_m\in \mathbb{Z}.
\end{equation}
 One should note that the discretization interval goes to zero in the 't Hooft limit $k\rightarrow \infty$, $N\rightarrow \infty$ with $k/N$ fixed. The saddle point equation for this partition function is
 \begin{eqnarray}
 V'(\ga_m)=\sum_{m\neq l}\cot{\ga_m-\ga_l\over 2}.
 \end{eqnarray}
 Now one can compare this situation with the thermal partition function of a Yang-Mills theory on $S^2\times S^1$. After integrating over all the massive modes, partition function for this case can be written as
 \begin{eqnarray}\label{eq:YMpart}
 \cZ_{YM}=\int_{-\infty}^\infty \prod_{m=1}^Nd\ga_m \left[\prod_{l\neq m}2 \sin\left( {\gt_l(\vec{n})-\gt_m(\vec{n})\over 2}\right)e^{-V_{YM}(U)}\right]
 \end{eqnarray}
 with $U$ being the zero mode of the holonomy around the thermal circle. This is a well studied version of unitary matrix model. The saddle point equation for this partition function is same as the previous one with summation replaced by integration. The repulsive term coming from the Haar measure always competes with the potential $V(U)_{YM}$. In the low temperature the repulsive term dominates and eigenvalues are distributed over the whole unit circle but in the high temperature limit $V_{YM}(U)$ becomes stronger  and confines the eigenvalues on some finite arc of the unit circle. This two distinct phases are termed as no-gap phase and one-gap phase.
 
 This same competition can be seen for the model (\ref{eq:fullpartition}) also, but in this case discretization of eigenvalues results into an upper limit of the eigenvalue density, independent of the exact form of $V(U)$. An interval of $\Delta\gt$ can contain maximum $\Delta n= \Delta\gt/(2\pi/k)$ eigenvalues. Eigenvalue density is defined as,
 \be
 \r(\q) = \lim\limits_{\Delta \q \ra 0} \frac1N{\Delta n\over \Delta \q }.
 \ee
 Hence, the maximum eigenvalue density can be $\rho(\gt)={k\over 2\pi N}={1\over 2\pi \lambda}$, where $\l=N/k$ ('t Hooft coupling) . Therefore $\rho(\gt)$ is bounded by
 \begin{eqnarray}
 \label{eq:evbound}
 0\le \rho(\theta)\le {1\over 2\pi \lambda}.
 \end{eqnarray}
This upper bound on eigenvalue density gives rise to the possibilities of new kinds of phase transitions in the theory. Generally the potential $V(U)$ are very complicated for CSM theories. \cite{shirazs2s1} has studied phase structure associated with different classes of CSM theories. Irrespective of their difference they all share same kind of phase structure. They all have a no-gap and a lower gap phase. Additionally they also have two new phases. The first one is the upper cap phase where there is no gap in eigenvalue distribution but it saturates its upper limit over some finite interval. The fourth phase consists of one upper cap and a lower gap. As a toy model, \cite{shirazs2s1} took Gross-Witten-Wadia model given by
\begin{eqnarray}\label{eq:GWWpf}
V(U)=-N \gb_1(\Tr U+\Tr U^{\dgr}).
\end{eqnarray}
GWW model has been studied by several authors. In large $N$ limit this model shows a third order phase transition from no-gap to one-gap. \cite{shirazs2s1} imposed upper bound on the GWW eigenvalue distribution and studied the enhanced phase structure. In the next two subsections we discuss their results. 

\subsection{Review of the capped GWW model}

Eigenvalue distribution function  $\rho(\gt)$ of capped matrix model can be decomposed into two parts
\begin{equation}
\rho(\gt)=\rho_0(\gt)+\psi(\gt),
\end{equation}
where $\rho_0(\gt)$ and $\psi(\gt)$ are continuous functions but with following boundary conditions
\begin{itemize} 
	\item $\rho_0(\gt)$ vanishes on lower gaps and equals to ${1\over 2\pi\gl}$ on the upper gaps,
	\item $\psi(\gt)$ vanishes on both the lower and upper gaps. 
\end{itemize}   
By doing this decomposition one converts the problem of solving the capped GWW matrix model to a usual matrix model problem with $\psi(\gt)$ being the eigenvalue distribution. This reparametrization effectively converts the upper caps to lower gaps and gives rise to a matrix model, where normalization of $\psi(\gt)$ depends on the choice of $\rho_0(\gt)$. One may also show that the solution is independent of choice of $\rho_0(\gt)$. Here we list four different phases for capped GWW matrix model obtained in \cite{shirazs2s1}.

\paragraph{No gap solution :} The no gap phase for capped GWW  model is identical with that of uncapped model. Eigenvalue distribution is given by
\begin{equation} \label{eq:evdistrinogap}
\rho(\gt)={1\over 2\pi}(1+2\gb_1\cos\gt).
\end{equation}
$\rho(\gt)$ is maximum (minimum) at $\gt=0\ (=\pi)$. Therefore from (\ref{eq:evbound}), we find no-gap phase is valid for
\ben\label{eq:nogapvalidity}
\begin{split}
\beta_1<{1\over 2\gl}-{1\over 2}\quad &\text{for } \gl>{1\over 2}\\
\beta_1<{1\over 2} \hspace{1.36cm} &\text{for }  \gl<{1\over 2}.
\end{split}
\een

\paragraph{Lower gap solution :} Eigenvalue distribution for this phase is also same as the one-gap solution for uncapped GWW model
\begin{equation}\label{eq:evdistrionegap}
\begin{split}
\rho(\gt)&={2\gb_1 \over \pi}\sqrt{{1\over 2\beta_1}-\sin^2{\gt\over 2}}\,\,\,\cos{\gt\over 2},\quad \text{for }\sin^2{\gt\over 2}<{1\over 2\gb_1}\\
\rho(\gt)&=0.\quad \text{for }\sin^2{\gt\over 2}>{1\over 2\gb_1}.
\end{split}
\end{equation}
The gap and distribution are distributed symmetrically around $\pi$. The maximum of this distribution is again at $\gt=0$. This phase only exists for $\gb_1\geq {1\over 2}$. Now we have further restriction due to upper limit of $\r(\q)$, which implies
\begin{eqnarray}\label{eq:onegapvalidity}
\gb_1\leq{1\over 8\gl^2}.
\end{eqnarray}  
Thus, lower-gap solution exists for 
\begin{equation}
	\gb_1<{1\over 8\gl^2} \quad \text{and} \quad \gl\leq\half{1}.
\end{equation}
For $\gl>\half{1}$ this solution does not exists.

\paragraph{Upper cap solution :} This is the first new phase in capped GWW matrix model as well as any capped matrix models. In this phase though eigenvalues are distributed like a no-gap solution but distribution is saturated over some finite range. Following \cite{shirazs2s1} one can find eigenvalue density for upper cap solution as 
\begin{equation}\label{eq:evdistrionecap}
\begin{split}
\rho(\gt)&={1\over 2\pi\gl}-2\gb_1{|\sin {\gt\over 2}|\over \pi}\sqrt{{{{1\over \gl}-1}\over 2\gb_1}-\cos^2{\gt\over 2}}\quad \text{for }\cos^2{\gt\over 2}<{{{1\over \gl}-1}\over 2\gb_1}\\
\rho(\gt)&={1\over 2\pi\gl}\hspace{5.5cm} \text{for }\cos^2{\gt\over 2}>{{{1\over \gl}-1}\over 2\gb_1}.
\end{split}
\end{equation}
The minimum of this solution occurs at $\gt=\pi$ with the value 
$${1\over 2\pi}\left({1\over \gl}-2\sqrt{2\gb_1}\sqrt{{1\over \gl}-1}\right).$$
Now apart from being real, minimum value should also be greater than zero. Hence, this solution exists for 
\begin{equation}
	{1\over 2\gl}-\half{1}<\gb_1<{1\over 8\gl(1-\gl)} \quad \text{for}  \quad \gl\geq{1\over 2}.
\end{equation}
Upper-cap solution does not exists for $\gl<\half{1}$.

\begin{figure}[h]
	\centering
	\includegraphics[width=11.0cm,height=7.5cm]{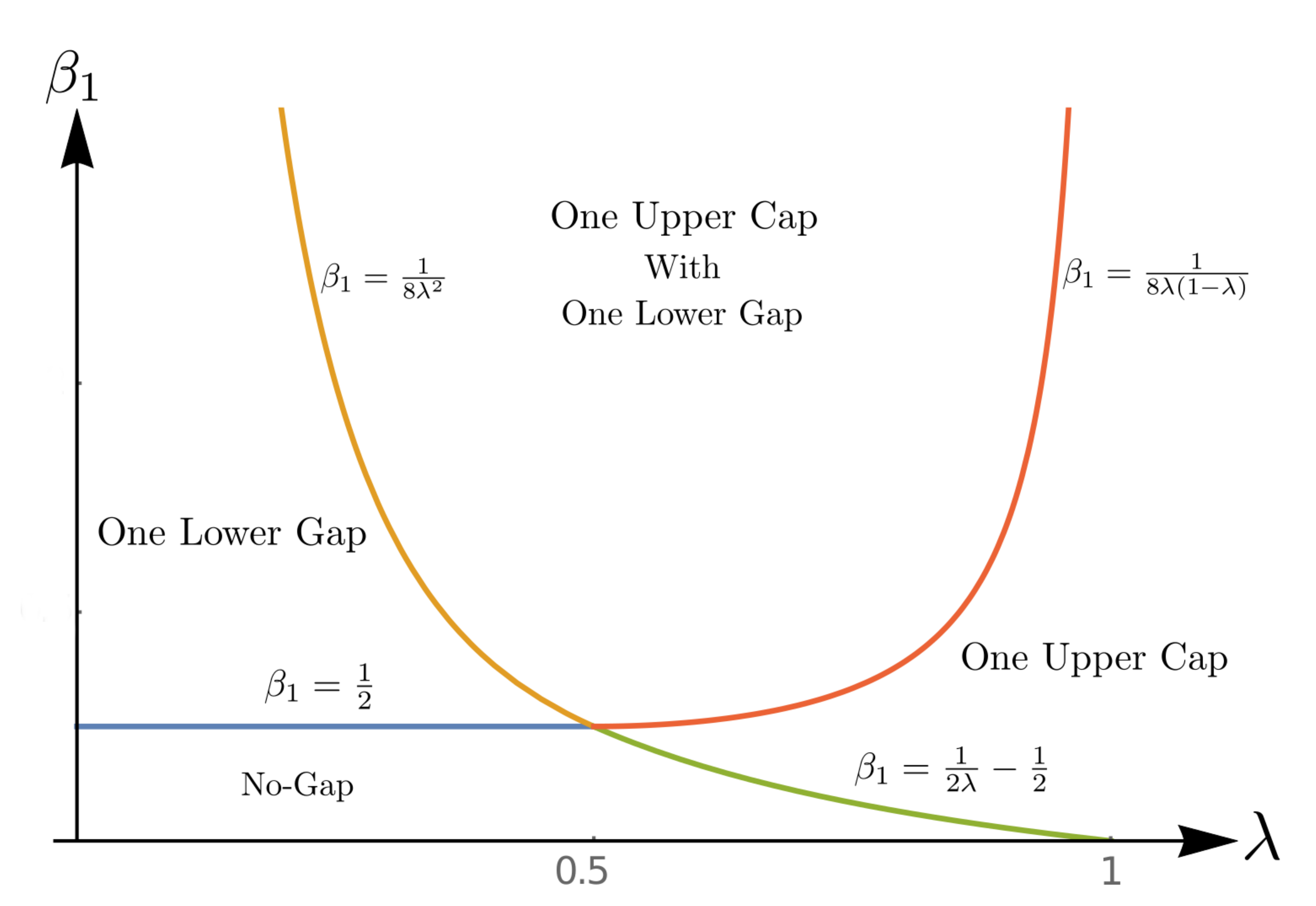}
	\caption{Phase Diagram for Capped GWW model}
	\label{fig:fullphase}
\end{figure}

\paragraph{Lower gap with upper cap :} Existence of lower gap solution also compels one to look for a solution which has one lower gap and one upper cap. If the upper cap is extended symmetrically around $\gt=0$ from $-a$ to $a$, and the lower gap extends from $-b$ to $b$ around $\gt=\pi$, then the eigenvalue distribution is given by
\begin{equation}\label{eq:evdistrioneloneu}
\begin{split}
\rho(\theta)&={|\sin {\gt}|\over 4\pi^2\lambda}\sqrt{(\sin^2{\gt\over 2}-\sin^2{a\over 2})(\sin^2{b\over 2}-\sin^2{\gt\over 2})}\,\,\,I(\gt)\quad\text{for }a\le|\gt|\le b\\
\rho(\theta)&={1\over 2\pi\gl} \hspace{7.5cm}\text{for }|\gt|\le a
\end{split}
\end{equation}
with
\begin{equation}
I(\gt)=\int_{-a}^{a}{d\ga\over(\cos\ga-\cos\gt)\sqrt{(\sin^2{a\over 2}-\sin^2{\ga\over 2})(\sin^2{b\over 2}-\sin^2{\ga\over 2})}}.
\end{equation}
Values of $a$ and $b$ can be deduced by solving the following constraint equations
\begin{eqnarray}
{1\over 4\pi\lambda}\int_{-a}^{a}{d\ga\over\sqrt{(\sin^2{a\over 2}-\sin^2{\ga\over 2})(\sin^2{b\over 2}-\sin^2{\ga\over 2})}}&=&2\beta_1,\\
{1\over 4\pi\lambda}\int_{-a}^{a}{\cos\ga \,\,d\ga\over\sqrt{(\sin^2{a\over 2}-\sin^2{\ga\over 2})(\sin^2{b\over 2}-\sin^2{\ga\over 2})}}&=&1+\gb_1(\cos a+\cos b).
\end{eqnarray}
Though it is a formidable task to compute this `two cut' distribution analytically, one can take some limits and check the range of validity for this solution. By taking the limits of $b\rightarrow\pi$ and $a\rightarrow 0$ one can deduce the range of validity of this solution as following
\ben
\begin{split}
	\gb_1 &\geq {1\over 8\gl(1-\gl)} \quad \text{for} \quad \gl\geq\half{1}\\
	\gb_1 &\geq {1\over 8\gl^2}\hspace{1.42cm} \text{for} \quad \gl\geq\half{1}.
\end{split}
\een
All the above discussions can be summarised in the following phase diagram give in figure \ref{fig:fullphase}.

\section{Phase space description of unitary matrix model}
\label{sec:youngreview}

Partition function of unitary matrix model can be written in different ways. The most well known form is given in terms of eigenvalues of unitary matrices $U$ (\ref{eq:YMpart}). Since the eigenvalues are like positions of free fermions, therefore 'eigenvalue representation' can
be thought of as position representations of partition function. The same partition function can also be expressed as a sum over representations of the unitary group \cite{douglas-kazakov,Kazakov:1995ae,duttagopakumar}. Therefore, in large $N$ limit, it is expected that the partition function is dominated by a particular representation. Representations of unitary group can also be cast in terms of Young tableaux. A dominant representation, therefore, implies a particular distribution of boxes in that diagram. This Young diagram representation of the partition function has a physical importance. The number of boxes in a row of a Young diagram behave like momentum of $N$ free fermions under consideration \cite{douglas2,Chattopadhyay}. Thus, Young representation is like a momentum representation of partition function. Young diagram distribution function captures information about the momentum distribution of $N$ fermions. A relation between position and momentum distribution functions allows one to provide a phase space description of different large $N$ phases of the unitary matrix model \cite{duttagopakumar, Chattopadhyay}.

The goal of this section is to find out the most dominant representations of a unitary matrix model in presence of GWW potential and obtain the underlying free fermionic phase space distribution for different large $N$ phases of the model in absence of any restriction on eigenvalue density. In the next section (sec. \ref{sec:constraint_rep}) we shall apply this method to obtain large $N$ representations for CS theory on \stso \ in presence of GWW potential and see how a restriction on eigenvalue distribution constrains Young distributions.

\subsection{Partition function in momentum basis}

We start with a generic partition function of the form 
\be\label{eq:partfuncplaq}
\cZ = \int [\cD U] \exp\lB N \sum_{n=1}^{\infty}{\beta_{n}\over n}
(\Tr U^{n}+\Tr U^{\dagger n})\rB,\quad \text{$\b_n$'s are some
	arbitrary coefficients.} 
\ee
GWW partition function (\ref{eq:GWWpf}) is a special case of this partition function :  $\b_{n}=0$ for $n\geq2$. We expand the exponential and write partition function as a sum over representations $R$ of unitary group $U(N)$,
\ben \label{eq:ZplaqsumoverR}
\cZ = \sum_R \sum_{\vec k} \frac{\varepsilon(\vec \b,\vec k)}
{z_{\vec k}} \sum_{\vec l} \frac{\varepsilon(\vec \b,\vec l)} {z_{\vec
		l}} \chi_{R}(C(\vec{k}))\chi_{R}(C(\vec{l})).  
\een
Here $\chi_{R}(C(\vec{k}))$ is the character of conjugacy class $C(\vec{k})$ of permutation group $S_{K}$, $K=\sum_n n k_n$ and
\ben
\varepsilon(\vec \b, \vec k) =
\prod_{n=1}^{\infty}N^{k_n}\b_{n}^{k_n}, \quad
z_{\vec k} = \prod_{n=1}^{\infty} k_{n}! n^{k_n}.
\een
To derive equation (\ref{eq:ZplaqsumoverR}) we have used the following identity
\be
\prod_n (\Tr U^n)^{k_n} = \sum_{R}\chi_{R}(C(\vec{k}))\Tr_{R}[U]
\ee
and the normalisation condition\footnote{See
	\cite{lasalle,hamermesh,fulton-harris} for details.}
\be \label{eq:normalization_TrU}
\int {\cal{D}}U\, \Tr_{R}[U] \Tr_{R'}[U^{\dagger}]=\delta_{RR'}.
\ee
Sum over representation of $U(N)$ can be written as a sum over different
Young diagrams. If $K$ is the total number of boxes in a Young diagram
with $n_i$ being the number of boxes in $i$-th row then
$\sum_{i=1}^N n_i=K$. Hence, sum over representations can be
decomposed as
\begin{equation}
\sum_R\longrightarrow\sum_{K=1}^\infty\,\sum_{\{n_i\}}\ 
\delta\lb \sum_{i=1}^Nn_i-K\rb \quad  \text{with}\quad 
n_1\geq n_2\geq\cdots\geq n_N\geq 0.
\end{equation}
The partition function, therefore, can be written as,
\ben\label{eq:pffinal} 
\cZ=\sum_{\vec n} \sum_{\vec k, \vec l}
\frac{\varepsilon(\vec \b, \vec k) \varepsilon(\vec \b, \vec
	l)}{z_{\vec k} z_{\vec l}} \chi_{\vec n}(C(\vec k)) \chi_{\vec
	n}(C(\vec l)) \ \delta \lb\sum_n n k_n-\sum_i n_i\rb \delta \lb
\sum_n n l_n-\sum_i n_i\rb .  
\een

We introduce $N$ variables $h_1, \cdots , h_N$, related to number of
boxes $n_i$'s as
\begin{eqnarray}\label{eq:h-nrelation}
h_i = n_i + N -i \qquad \forall \quad i =1, \cdots, N.
\end{eqnarray}
$h_i$'s are shifted number of boxes\footnote{Our terminology is little
  sloppy. We also call $h_i$ as the number of boxes in the $i$-th
  row.}. From monotonicity of $n_i$s it follows that $h_i$s satisfy
the following constraint
\be
\label{hmonotonicity} h_1> h_2> \cdots > h_N \geq 0.  
\ee
From now on we shall use variables $h_i$s in stead of $n_i$s.

In large $N$ limit we define continuous variables
\begin{eqnarray}\label{eq:contvardef}
{h_{i}\over N}=h(x), 
\quad\quad k_{n}=N^{2}k'_n,
\quad\quad
x={i\over N} \quad \with \quad x\in [0,1].
\end{eqnarray}
In large $N$ limit, dominant contribution comes from diagrams with box numbers $\sim \cO(N^2)$, hence we separate out that factor and all $k_n'$s are $\cO(1)$ number. Also, in this limit summation over $i$ is replaced by an integral over $x$,
\be
\sum_{\a=1}^{\infty} \ra N\int_0^1 dx
\ee
and sum over representations ($\vec n$) and sum over cycles ($k_n$)
are given by path integral over $h(x)$ and integral over
$k_n'$. Writing characters $\chi_{\vec h}$ in terms of $h(x)$ and
$k_n'$, partition function (\ref{eq:pffinal}), in large $N$ limit, can
be written as
\be\label{eq:pfyt1}
\cZ = \int [\cD h(x)] \prod_n\int dk_n' dl_n' \exp\lB -N^2 
S_{\text{eff}} [h(x),\vec {k_n'},\vec{l_n'}]\rB,
\ee
where $S_{\text{eff}}$ is the effective action. Thus we see how one can write down the partition function in Young tableaux basis (momentum basis). Dominant contribution
to partition function comes from those representations which maximise effective action $S_{\text{eff}}$. 

For the simplest case (GWW), $\b_1 \neq 0$ and other $\b_n=0$ (for
$n\geq 2$) one can find $u(h)$ for different phases of the model
\cite{duttagopakumar, duttadutta}. In this simple case we need to
calculate character of permutation group in presence of one cycles
only ($k_n=0$ for $n\ge 2$) which is given by dimension of the
corresponding representation
\be
\chi_{\vec h}(k_1) = \frac{k_1!}{h_1!\ h_2! \cdots h_N!}\prod_{i<j}(h_i-h_j).
\ee
Plugging this expression in equation (\ref{eq:pfyt1}) we find,
\be\label{eq:pfGWW2}
\cZ = \int [\cD h(x)] \exp\lB -N^2 
S_{\text{eff}} [h(x),{k_1'}]\rB
\ee
where,
\be \label{eq:effacnGWW}
-S_{\text{eff}} [h(x),{k_1'}] = \int_0^1 dx \Xint-_0^1 dy \ln |h(x)-h(y)| -2 \int_0^1 dx \ h(x) \ln h(x) + 2 k_1' \ln \bo +2k_1' +1
\ee
with
\be
k_1' = \int_0^1 dx \ h(x) - \half1. 
\ee 

\subsection{The saddle point equation}

Now we can carry out a saddle point analysis for the effective action (\ref{eq:effacnGWW}). Varying $\seff{h(x),k_1}$ with respect to $h(x)$, we obtain the saddle point equation
\ben\label{eq:sadeqn-h}
\Xint-_0^1 dy {1\over h(x) - h(y)} = \ln \lB{h(x) \over \bo}\rB .
\een 
We introduce a function called Young tableaux density defined as,
\be \label{eq:defuh}
u(h) = - {\dow x\over \dow h}.
\ee
$u(h)$ captures information how boxes are arranged in a Young diagram. Different Young diagrams correspond to different $u(h)$.  Since $h(x)$ is a monotonically decreasing function of $x$, Young density has a lower bound $u(h) \ge 0 \ \forall \ x\in[0,1]$ with a normalization condition (follows from the definition)
\be
\int_{h_L}^{h_U} dh \ u(h) =1,
\ee
where the interval of support $[h_L,h_U]$ is specified by $h_L = h(1)$ and $h_U=h(0)$. Also, from equation (\ref{eq:h-nrelation}) we find,
\be
{\dow n(x)\over \dow x} = 1-{1\over u(h)}.
\ee
Since $n(x)$ is monotonically decreasing function therefore, ${\dow n(x)\over \dow x} \leq 0$. Hence, $u(h)$ also satisfies an upper bound 
\be\label{eq:constraintuh}
u(h)\leq 1.
\ee

In terms of Young density $u(h)$ the saddle point equation (\ref{eq:sadeqn-h}) can also be written as
\be\label{eq:saddlepointYT}
\Xint-_{h_L}^{h_U} dh' {u(h')\over h-h'} = \ln\lB{h \over \bo}\rB.
\ee
Given a distribution $u(h)$ total number of boxes in that representation is given by
\be
k_1 = N^2 k_1' = N^2 \int_{h_L}^{h_U} dh \ h  u(h) -\half{N^2}.
\ee
Therefore, one needs to solve this saddle point equation in presence of the constraint on $u(h)$ (\ref{eq:constraintuh}). The saddle equation admits different classes of solutions depending whether $u(h)$ saturates the upper bound or not. A thorough study of this model has been done in \cite{duttagopakumar,duttadutta}. There exists a phase transition between these saddle points as one varies the parameter $\bo$. This is similar to the large $N$ phase transition of Douglas-Kazakov \cite{douglas-kazakov} in $2d$ Yang-Mills theory.

\subsection{Eigenvalue distribution vs. Young tableaux distribution}

Before we delineate how constraint on eigenvalue density restricts Young distribution, we need to understand how these two distribution functions (eigenvalue and Young diagram) are related to each other for different large $N$ phases for GWW matrix model \cite{gross-witten,wadia} when there is no restriction on eigenvalue density \cite{duttagopakumar}. As mentioned in the last section, the partition function/action (\ref{eq:GWWpf}) has two possible phases in large $N$ limit, when there is no restriction on eigenvalue density: (i) no-gap phase, eigenvalue distribution is given by (\ref{eq:evdistrinogap}) for $\b_1< \half1$, (ii) one-gap solution, eigenvalue distribution is given by equation (\ref{eq:evdistrionegap}) for $\b_1 > 1/2$. There exists a third order phase transition at $\b_1 =1/2$, known as Gross-Witten-Wadia phase transition.

\begin{figure}[h]
	\centering
	\begin{subfigure}{0.4\textwidth}
		\centering
		\includegraphics[width=6cm,height=5cm]{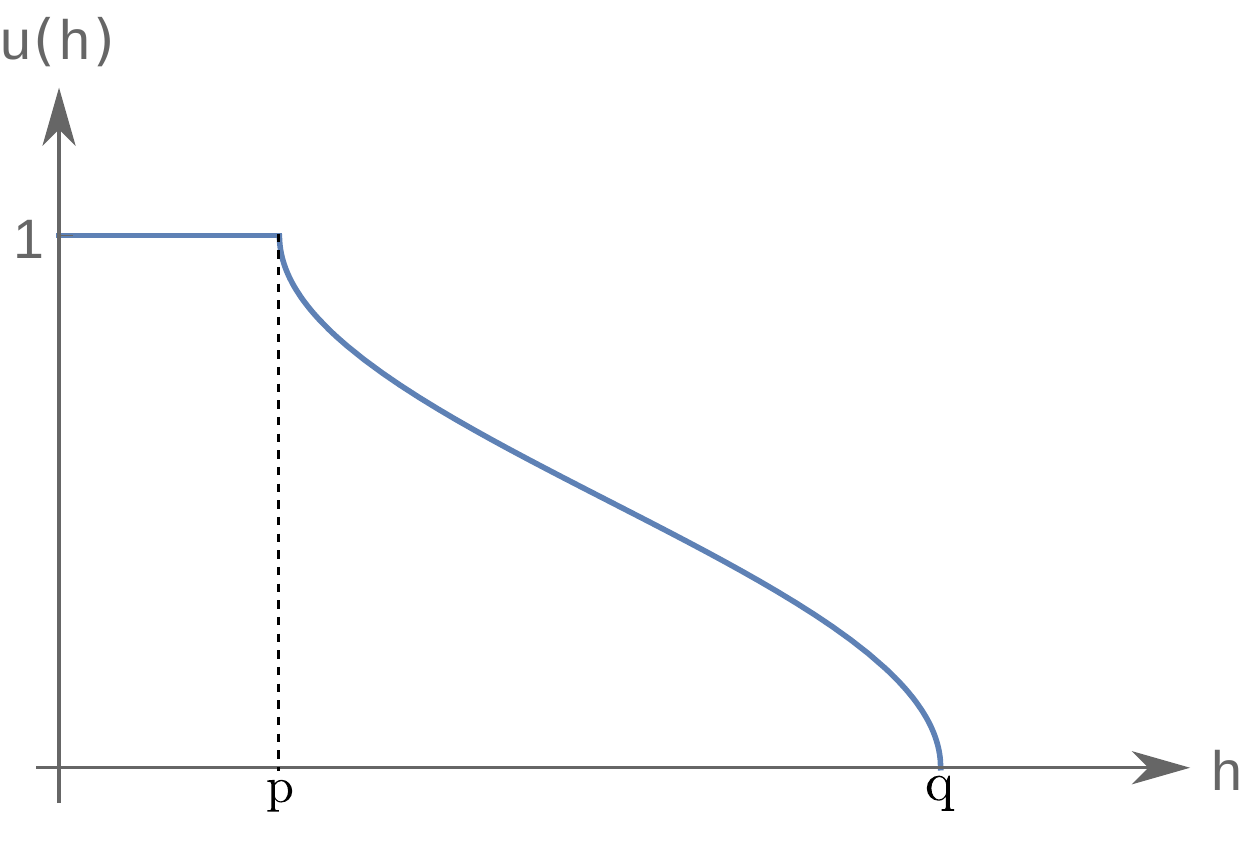}
		\caption{$u(h)$ vs. $h$ for no-gap phase. $u(h)$ has a continuous support between $0$ and $q=1-2\bo$.}
	\end{subfigure}%
	\hspace{.8cm}
	\begin{subfigure}{0.4\textwidth}
		\centering
		\includegraphics[width=6cm,height=5cm]{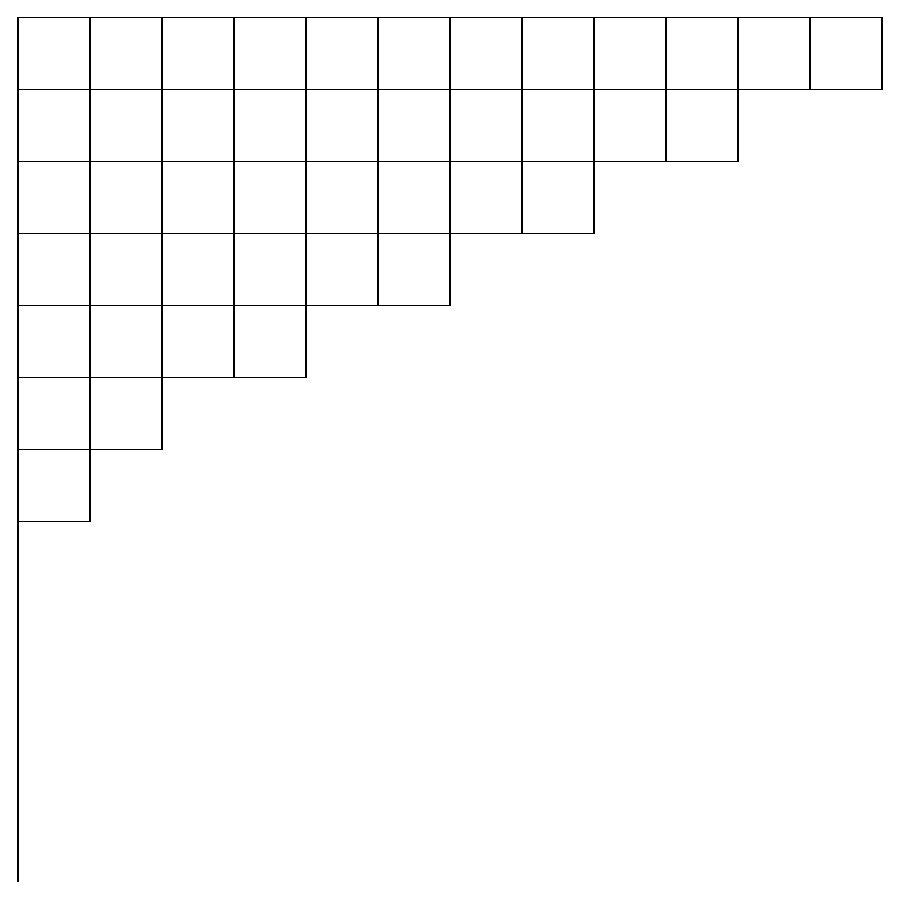}
		\caption{A typical Young diagram for no-gap phase. Here we see that a finite number of rows are empty.}
	\end{subfigure}
	\caption{Young distribution and Young diagram for no-gap phase.}
	\label{fig:uhnogap}
\end{figure}

In Young tableaux side, one needs to solve the saddle point equation (\ref{eq:sadeqn-h}) in presence of the constraint $0\leq u(h)\leq 1$. The equation admits two possible classes of solutions. In the first class, $u(h)$ saturates the maximum value in a finite range of $h$
\ben\label{eq:uhnogap}
\begin{split} 
	u(h) &= 1, \hspace{3cm} 0\leq h \leq p \\
	& = \frac{1}{\pi} \cos^{-1} \lB {h-1 \over 2\b_1}\rB, \quad p\leq h\leq q
\end{split}
\een
where
\be
p = 1-2\b_1, \quad q= 1+2\b_1. 
\ee
This class of solution exists for $\b_1\le 1/2$. Young tableaux distribution $u(h)$ vs. $h$ and a typical Young diagram has been depicted in figure \ref{fig:uhnogap}. 
%%%%%%%%%%%%%%%%%%%%%%%%%%%%%%%%%%%%%%%%%%%%%%%%%%%%%
%%%%%%%%%%%%%%%%%%%%%%%%%%%%%%%%%%%%%%%%%%%%%%%%%%
%%%%%%%%%%%%%%%%%%%%%%%%%%%%%%%%%%%%%%%%%%%%%%%%%%%%%
%%%%%%%%%%%%%%%%%%%%%%%%%%%%%%%%%%%%%%%%%%%%%%%%%%
Free energy computed for this representation matches with free energy of non-gap phase in eigenvalue side \cite{duttagopakumar}. Hence, we identify this phase with no-gap phase. 

In the second class ($\b_1>1/2$), $u(h)$ never saturates the upper bound
\ben\label{eq:uhonegap}
u(h) = \frac2\pi \cos^{-1} \lB {h+\b_1 -1/2 \over 2\sqrt{\b_1  h}}\rB, \quad p\leq h\leq q
\een
where,
\be
p = \lb \sqrt{\b_1} -{1\over \sqrt2}\rb^2, \quad q = \lb \sqrt{\b_1} +{1\over \sqrt2}\rb^2 .
\ee
\begin{figure}[h]
\centering
\includegraphics[width=10cm,height=5cm]{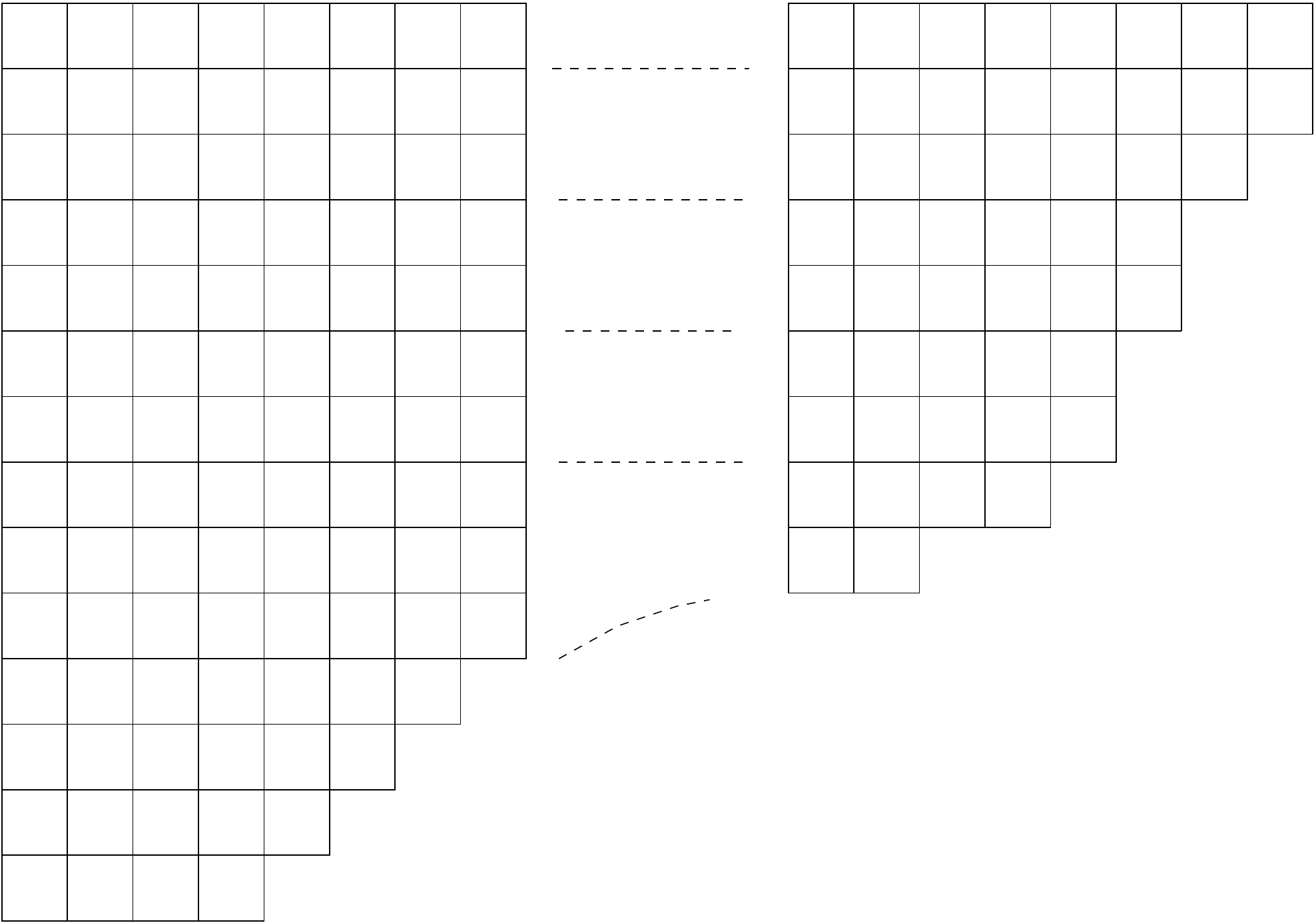}
	\caption{A typical Young diagram for one-gap phase}
	\label{fig:lowergap}
\end{figure}
Free energy for this branch matches with free energy of one-gap solution in eigenvalue side. Hence, this representation is complimentary to one-gap eigenvalue representation.

\subsubsection*{Relation between eigenvalue and Young diagram distributions}

It was first observed in \cite{duttagopakumar} that there exists an one to one correspondence between Young distribution and eigenvalue distribution for different phases at large $N$. Namely, they are functional inverse of each other. For no gap phase $\b_1<1/2$, it is given by (from equation (\ref{eq:evdistrinogap} and \ref{eq:uhnogap}))
\ben
\label{eq:phspident1}
\r(\q) = {h\over 2\pi}, \qquad u(h) = {\q \over \pi}.
\een
Here we see that eigenvalue density is identified with $h$ (hook numbers) divided by $2\pi$. Therefore, maximum value of eigenvalue density is given by maximum spread (or support) of $u(h)$. In this case maximum spread (support) is $q=1+2\b_1$.

\begin{figure}[H]
	\centering
	\begin{subfigure}{0.4\textwidth}
		\centering
		\includegraphics[width=6cm,height=5cm]{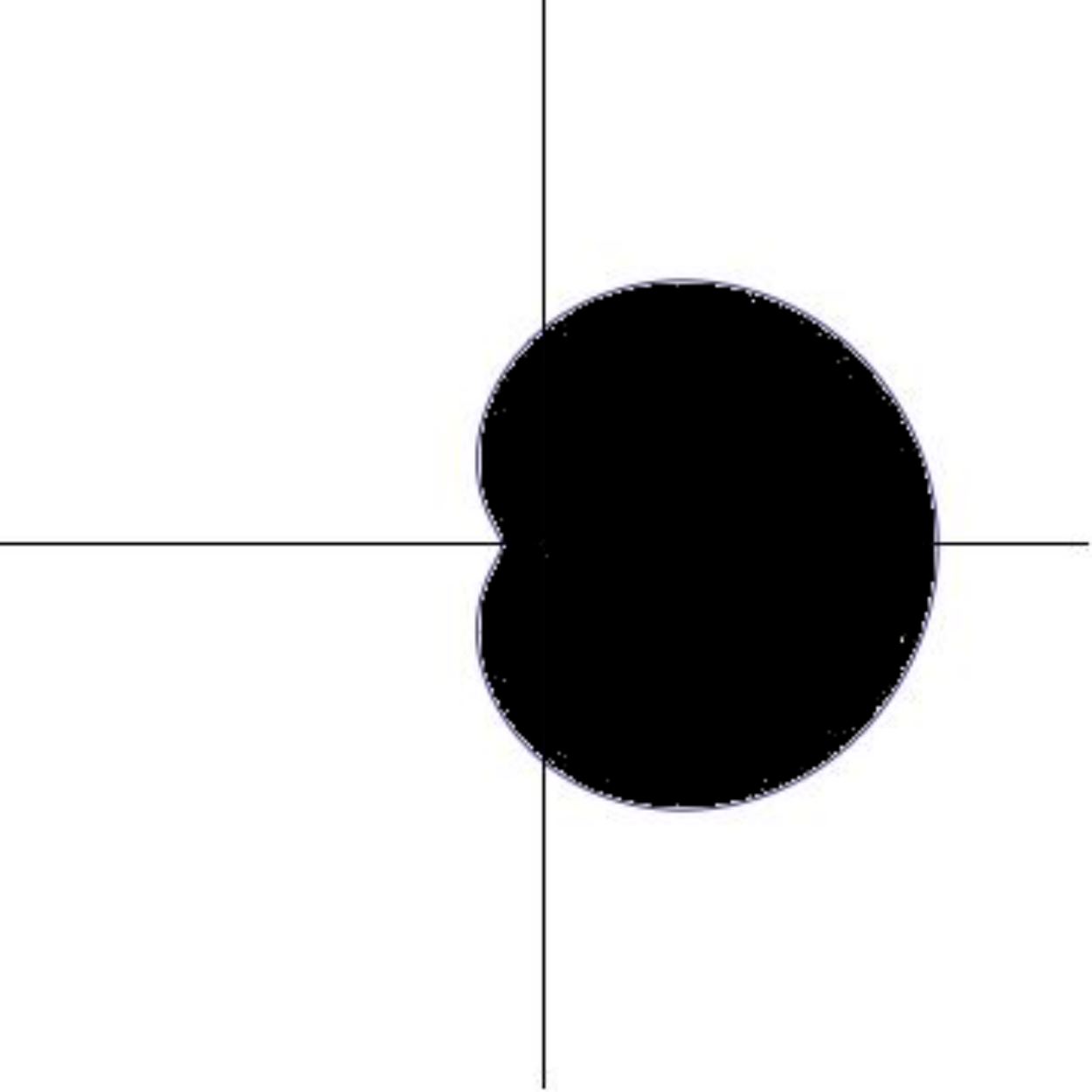}
		\caption{Phase space distribution for no-gap phase.}
	\end{subfigure}%
	\hspace{.8cm}
	\begin{subfigure}{0.4\textwidth}
		\centering
		\includegraphics[width=6cm,height=5cm]{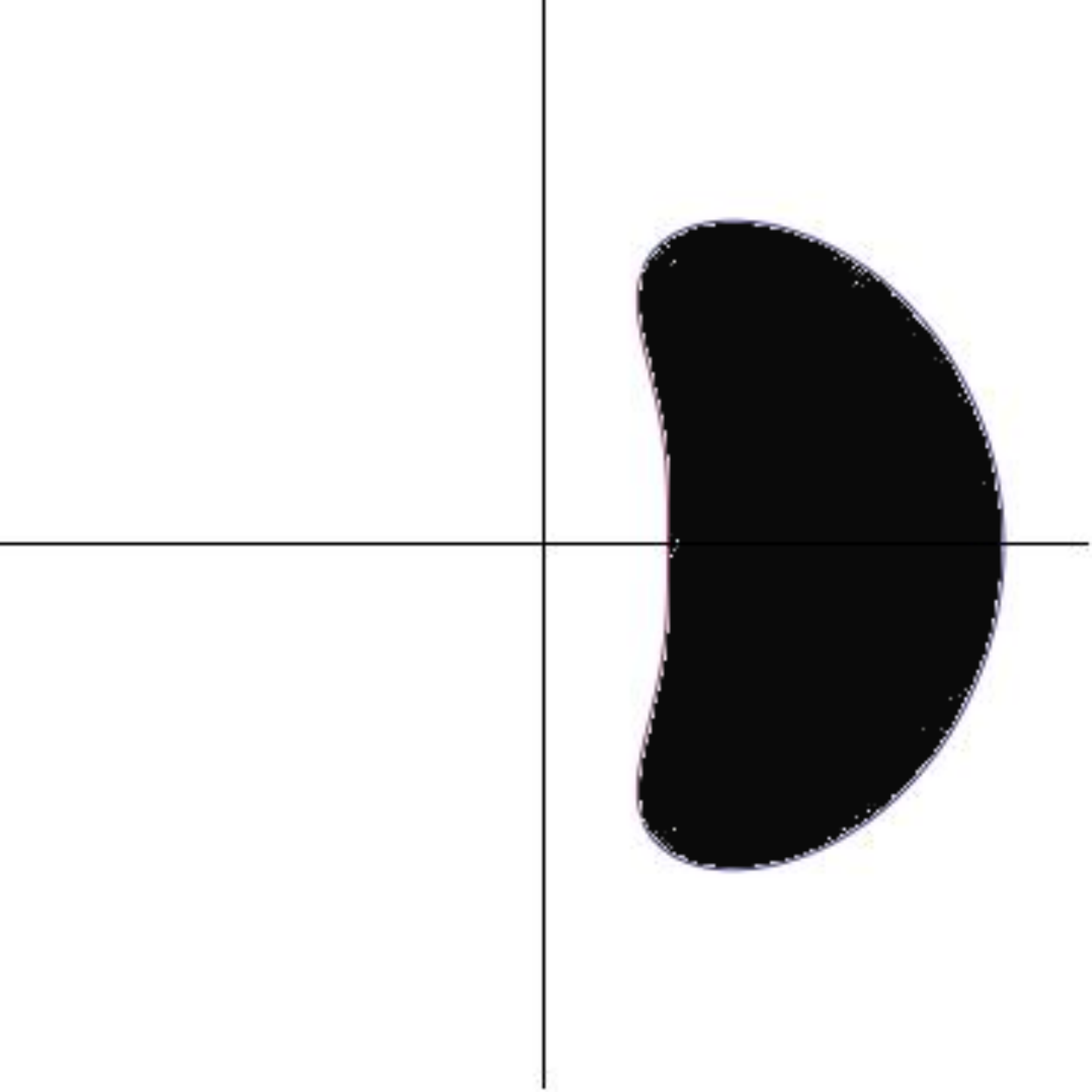}
		\caption{Phase space distribution for one-gap phase.}
	\end{subfigure}
	\caption{Phase space distribution for generic GWW matrix model.}
	\label{fig:phasespacenoonegap}
\end{figure}

For one-gap phase, the relations are given by
\be
\label{eq:phspident2}
\r(\q)={h_+(\q)-h_-(\q)\over 2\pi}, \quad u(h) = {\q \over \pi}
\ee
where, $h_+(\q)$ and $h_-(\q)$ are the solutions of
\be
h^2-\lB 1+2\bo \cos(\pi u(h))\rB h + \lb \bo -\half1\rb^2 =0, 
\ee
which is obtained from (\ref{eq:uhonegap}). Again we see that the maximum value of eigenvalue density is given by the maximum spread of $u(h)$. We also note that in this simple model identification between $u(h)$ and $\q$ is simple and same in both the phases. However, for a generic model this may not be the case \cite{riemannzero,Chattopadhyay}. The above identifications provide an emergent phase space description of unitary matrix model. The phase space distribution is similar to Thomas-Fermi distribution of some free fermions moving on $S^1$. Eigenvalues ($\q_i$s) are coordinates of these fermions and hook numbers ($h_i$s) are momentum of these fermions. The above identifications defines branches of fermi surfaces in phases space. One can define a phase space density $\o(h,\q)$ in $\lb h,\q\rb$ plane with the following properties (we consider $h$ as radial coordinate),
\be
\begin{split}
\o(h,\q) = {1\over 2\pi}\Theta\lB(h_+(\q)-h)(h-h_-(\q)) \rB
\end{split}
\ee
such that
\be\label{eq:rhoufromps}
\r(\q)= \int_0^\infty \o(h,\q)dh, \qquad u(h)= \int_{-\pi}^{\pi} \o(h,\q)d\q.
\ee
Phase space distributions for no-gap and one-gap phases have been plotted in figure \ref{fig:phasespacenoonegap}.
%%%%%%%%%%%%%%%%%%%%%%%%%%%%%%%%%%%%%%%%%%%%%%%%%%%%%
%%%%%%%%%%%%%%%%%%%%%%%%%%%%%%%%%%%%%%%%%%%%%%%%%%
%%%%%%%%%%%%%%%%%%%%%%%%%%%%%%%%%%%%%%%%%%%%%%%%%%%%%
%%%%%%%%%%%%%%%%%%%%%%%%%%%%%%%%%%%%%%%%%%%%%%%%%%

\subsection{Reduced Young diagram}
\label{sec:reducedYD}

\begin{figure}[H]
\centering
\includegraphics[width=9cm,height=5cm]{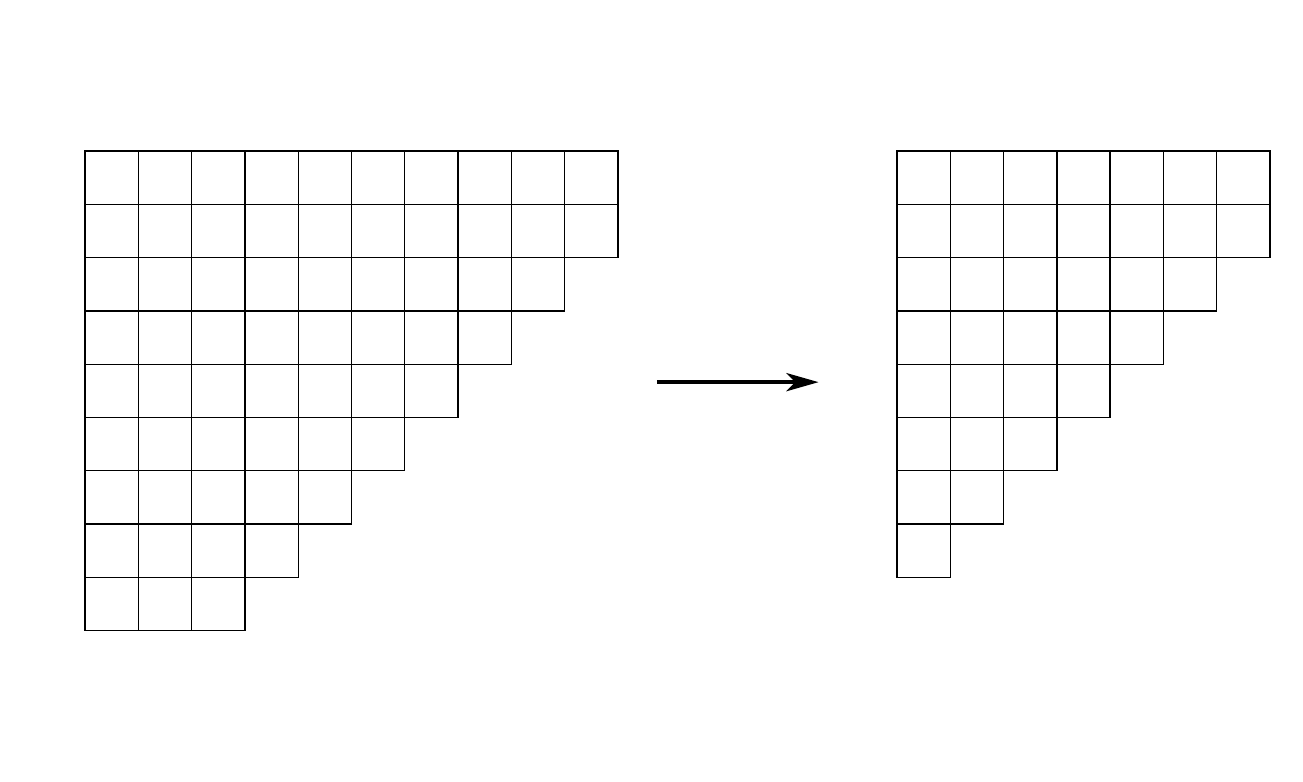}
	\caption{Equivalent Young diagrams.}
	\label{fig:YTeqv}
\end{figure}

The Young diagram \ref{fig:lowergap} for lower-gap phase, obtained by solving saddle point equation, is equivalent to a diagram with all $p$ columns omitted from the left, as shown in figure \ref{fig:YTeqv}.

The diagram on the right is obtained by taking conjugation of conjugation of the first diagram. Since conjugation of conjugation of any representation is the representation itself, hence both the diagrams in figure \ref{fig:YTeqv} describe the same representation.
Therefore, one can define a conjugacy class of Young diagrams. Two diagrams $Y_1$ and $Y_2$ belong to the same conjugacy class if $(Y_1^{\dagger})^{\dagger}=Y_2$. It is, therefore, obvious that all the diagrams belong to the same conjugacy class will have the same shape or pattern of the edge i.e. $u(h)$. Hence, the shape $u(h)$ represents a conjugacy class. The diagram on the right in figure \ref{fig:YTeqv} is called {\it reduced Young diagram}. From now and onwards, whenever we talk about Young diagrams we always mean reduced Young diagrams.

\section{Constraint on Young diagrams for Chern-Simons-Matter theory on $S^2\times S^1$}
\label{sec:constraint_rep}

So far, we have discussed about a generic GWW matrix model. CS matter theory with GWW potential have same matrix model in large $k$, large $N$ limit but with an extra condition on eigenvalue density $\r(\q)\leq {1\over 2\pi \l}$. We use the above identification between Young distribution and eigenvalue distribution to impose bound on dominant Young distributions for CS-GWW-mater theory on $S^2\times S^1$.

\subsection*{Our claim}
We see that the eigenvalue density is related to spread (support) of $u(h)$ i.e. 
width of Young distribution. Thus an upper cap on eigenvalue density puts 
restriction on spread of $u(h)$ or width of Young diagrams. Since, spread of Young 
distribution is identified with $2\pi \r({\q})$, we {\it claim} that
%\begin{tcolorbox}
\begin{center}
		\setlength{\fboxrule}{1.5pt}
\fbox{\begin{minipage}{0.9\textwidth}
for CS-matter theory on $S^2\times S^1$ the dominant representations have a
Young distribution function with maximum spread $1/\l$.
\end{minipage}}
\end{center}
%\end{tcolorbox}
\vspace{0.2cm}
Imposing this extra condition on distribution function one has to solve saddle point equation (\ref{eq:sadeqn-h}) to find dominant representations. The solutions to this equation, which we have already obtained, are also going to be the solution of CS-GWW-matter theory with the restriction as mentioned in our claim. However, there could be new class of solutions with this extra condition on Young distribution function. In this section we try to obtain those new class of solutions and find corresponding phase space distribution.

Before we find new class of solutions, we first check whether our claim is consistent with the region of validity of existing solutions for no-gap and lower-gap phases. We see that for $\bo < 1/2$ spread of $u(h)$ is given by $q=1+2\bo$. Imposing an upper bound on q ($q\leq {1\over \l}$), we find that this solution is valid for
\be
\qquad \bo \leq {1\over 2\l} -\half1.
\ee
Thus we see that no-gap phase is valid for $\bo\leq 1/2$ and $\bo \leq {1\over 2\l} -\half1$, which exactly matches with validity of no-gap solution for CS-GWW-matter theory (\ref{eq:nogapvalidity}).

For lower-gap phase, the spread of $u(h)$ is given by $q-p = 2\sqrt{2\bo}$, which is equivalent to the width of reduced Young diagram. Imposing the condition on spread we find that this phase is valid for
\be
\bo\geq \half1, \quad \bo\leq {1\over 8\l^2},
\ee
which is similar to (\ref{eq:onegapvalidity}).

\subsection{Integrable representations}

An integrable representation of affine Lie algebra $SU(N)_k$ is characterised by an Young diagram which has maximum $k-N$ boxes in the first row. Using the relation between number of boxes $n_i$ and hook length $h_i$ : $h_i=n_i+N-i$, we see that integrable representations correspond to diagrams with
\be
h_1 < k \quad \text{which implies} \quad h(0) <\frac1\l, \quad (h(0)=h_1/N).
\ee

For no-gap phase we have $h(0)=q$ and since $q<\frac1\l$ we conclude that no-gap phase of this theory belongs to integrable representation of $SU(N)_k$ affine Lie algebra. Similarly, for lower-gap phase we consider reduced Young diagrams (defined in section \ref{sec:reducedYD}). The reduced diagrams have number of boxes in the first row less than $1/\l$. Hence, lower-gap phase also belongs to integrable representations of $SU(N)_k$ affine Lie algebra. Thus from phase space identification (\ref{eq:phspident1} and \ref{eq:phspident2}) we see that our claim in turn implies that putting a cap on eigenvalue distribution constraints the corresponding representations to be integrable. In the next section we shall see that this is true for upper-cap phase as well.

It was explicitly shown in \cite{Chattopadhyay} that eigenvalues $\q_i$'s and number of boxes $h_i$ behave like position and momentum of $N$ non-interacting fermions moving on $S^1$. The Hamiltonian for this system can also be written as real bosonic field theory with collective field $\r(\q,t)$ on $S^1$ on a fixed time slice \cite{das-jevicki,jevicki,sakita}. For CS matter theory on $S^1$ the coordinates of free fermions i.e. eigenvalues are discrete with minimum spacing $2\pi/k$. This implies that corresponding reciprocal space ($h$ space) is periodic and the size of first Brillouin zone is $k$ ($h_{\text{max}}=k$) which is consistent with our claim. Also, since position space is periodic with period $2\pi$ the reciprocal $h$ space is discrete with minimum spacing $1$.

\subsection{Finding representation for upper-cap phase}

\begin{figure}[h]
	\centering
		\includegraphics[width=16cm,height=8cm]{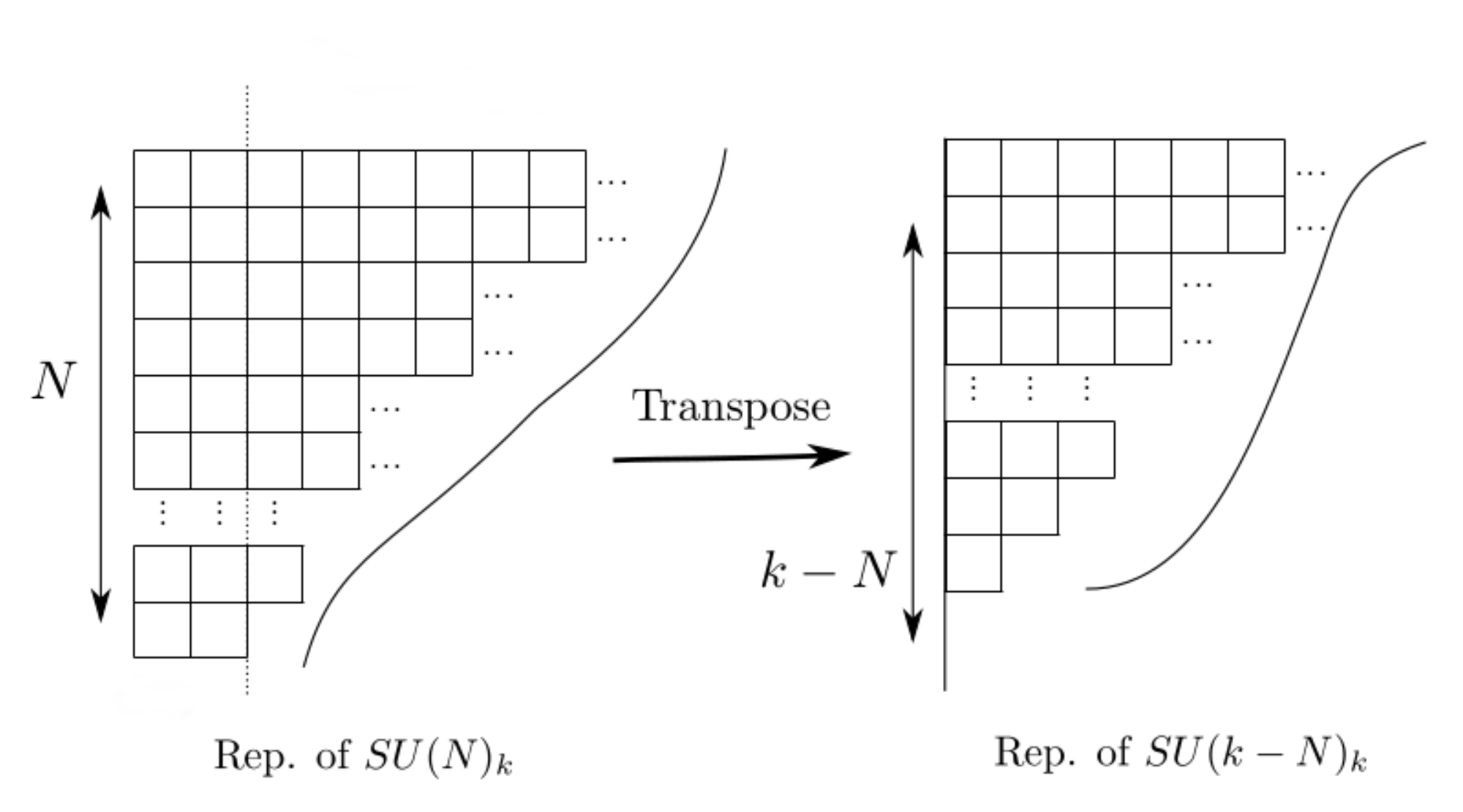}
	\caption{Level-rank duality and transposition of Young diagrams.}
	\label{fig:motivate}
\end{figure}

Using our claim on $u(h)$, one can find the dominant representation for the upper cap phase\footnote{To construct a new phase, without using level-rank duality, we start from a point in phase diagram (figure \ref{fig:fullphase}) in no-gap phase ($\bo\leq 1/2$ and $\bo \leq {1\over 2\l} -\half1$) and increase the value of $\l$ keeping $\bo$ fixed. The spread of $u(h)$ will remain constant as we are not changing $\bo$. However, the cut-off value decreases as we increase $\l$. Therefore, the spread of $u(h)$ will saturate the upper value (i.e. $1/\l$) when we touch the transition line. After that the boxes in Young diagram will reorganise themselves as we cross the transition line. The only possible way to reorganise the shape to maintain the constraint is to increase the number of boxes at $x=p$ such that spread of $u(h)$ satisfies the upper bound. A typical distribution is given by equation (\ref{eq:uhupcapansatz})}. However, here we use the level rank duality to find the dominant representation for the upper cap, and verify our claim. It has been argued in \cite{Naculich:1990} that under level-rank duality the primary fields in the dual theory will transform under the integrable representation $\tilde{Y}$ of $SU(k_{YM})_N$. Where $Y$ and $\tilde{Y}$ is related by ``transposition", i.e the Young diagrams for these representations are related by the interchange of rows and columns. We know that Lower gap is mapped to upper cap via level rank duality \cite{shirazs2s1}. Therefore, the dominant representations for upper-cap phase can be obtained by transposing the dominant representations for the lower-gap phase. See figure \ref{fig:motivate}.

\begin{figure}[h]
	\centering
	\includegraphics[width=7cm,height=6cm]{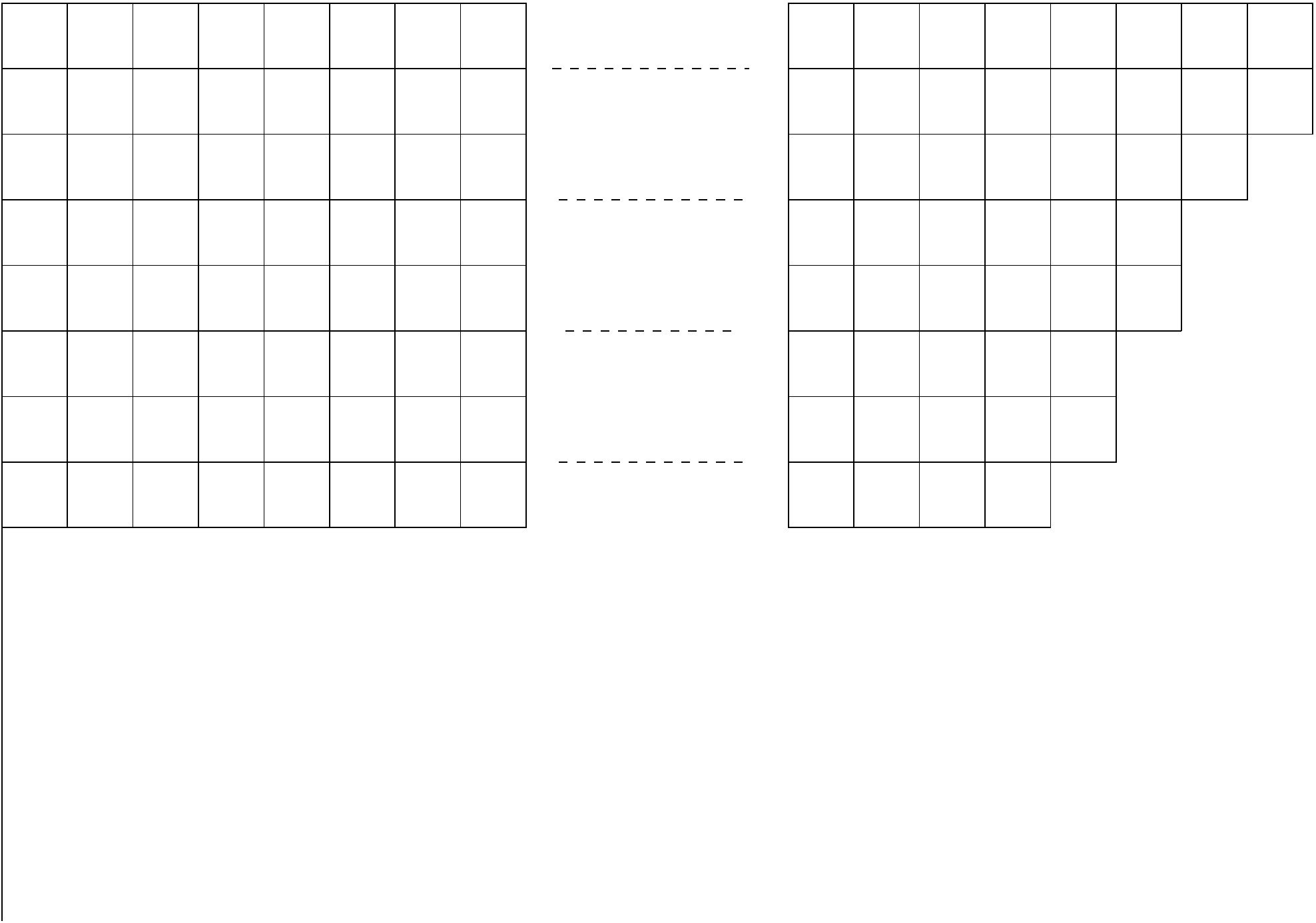}
	\caption{A typical Young diagram for upper-cap phase.}
	\label{fig:YDuppercap}
\end{figure}

We start with the reduced Young diagram for lower-gap phase. The diagram has $N$ number of rows and the first row has number of boxes less that $k-N$. Therefore, when we transpose the diagram we get a new diagram with number of rows less than $k-N$. This indicates that the transposed diagram is a representation of $SU(k-N)$. Also the first row of transposed diagram has maximum $N$ number of boxes. Therefore the diagram represents an integrable representation of $SU(k-N)_k$. This observation prompts us to take the following generic ansatz for upper-cap phase.
\ben\label{eq:uhupcapansatz}
\begin{split}
u(h) &= 1\hspace{1cm} 0< h< p \\
&= 0 \hspace{1cm} p<h<q \\
&= \tilde{u}(h)\quad q<h<r
\end{split}
\een
with 
\be
\label{eq:pvalueuppcap}
p=2-{1\over \l}.
\ee
 The choice of $p$ follows from the fact that when $\l \ra 1/(1+2\bo)$ for a fixed $\bo<1/2$, we should get back the Young diagram for no-gap phase. According to our claim we also have $r-s\leq 1/\l$. This ansatz corresponds to a Young diagram with no boxes from $x=1$ to $x=1-p=1/\l-1$. Then there is a sudden jump (of the order of $N$) in number of boxes from {\it zero} to $q-1+2\bo$. After that number of boxes (in each row) continuously increases to $r-1$. A typical Young diagram for this distribution is drawn in figure \ref{fig:YDuppercap}.

\begin{figure}[h]
	\centering
	\begin{subfigure}{0.4\textwidth}
		\centering
		\includegraphics[width=10cm,height=6cm]{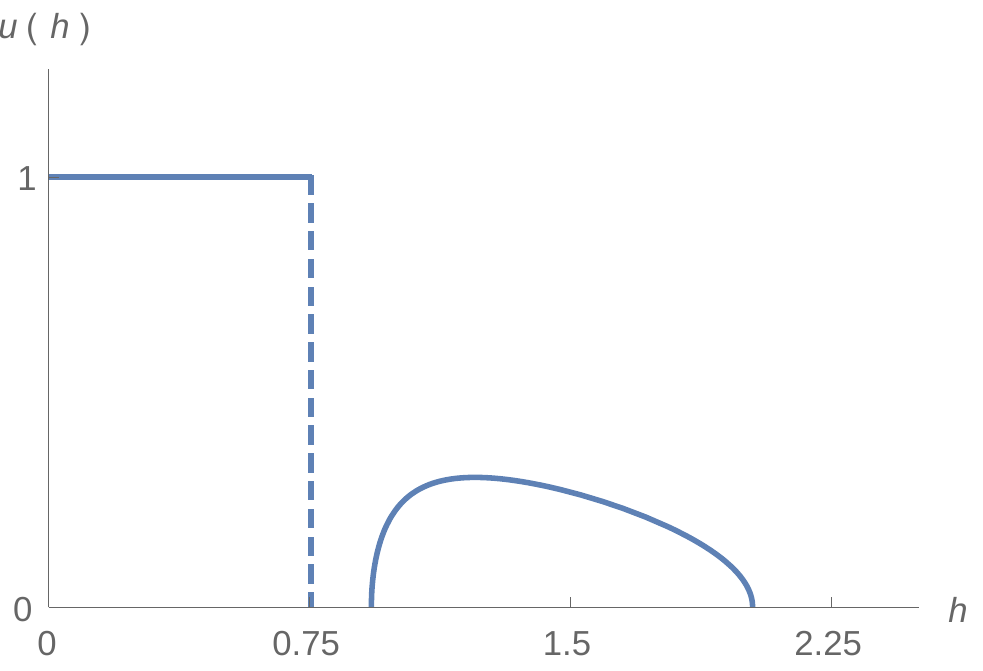}
%		\caption{Phase space distribution for no-gap phase.}
	\end{subfigure}%
	\caption{$u(h)$ vs. $h$ for upper-cap solution.}
	\label{fig:uhupcap}
\end{figure}

$u(h)$ being normalised to unity, we have
\begin{equation}
\int_q^r \tilde{u}(h)\,dh=1-p\quad \text{and}\quad
\int_q^r h\,\tilde{u}(h)\, dh=k'+\half{1}+\half{p^2}.
\end{equation}
The saddle point equation (\ref{eq:saddlepointYT}) for $\tilde{u}(h)$ is given by 
\begin{equation}
\Xint-_q^r{\tilde u(h')\over h-h'}dh'=\lnb{h-p\over \bo}.
\end{equation}
To solve this equation we define a resolvent
\begin{equation}
\label{eq:resolventH}
H(h)=\int_q^r{\tilde u(h')\over h-h'}dh'
\end{equation}
with the following defining properties
\begin{enumerate}[label=(\alph{enumi})]
	\item $H(h)$ is an analytic function of $h$ except on the support, where it has a branch cut (in complex $h$ plane).
	\item $H(h)$ is real for real and positive $h$ outside the support.
	\item Asymptotically $H(h)$ is given by
	\begin{equation}
	H(h\rightarrow \infty)\sim {1-p\over h}+{1\over h^2}\left(k'+\half{1}+\half{p^2}\right)+O({h^{-3}}).
	\end{equation}
	\item $\displaystyle{\lim_{\gep\rightarrow 0}}\,\left[H(h+i\gep)+H(h+i\gep)\right]=2\lnb{h-p\over \bo}$ for $h\in(q,s)$.
	\item Young distribution $u(h)$ is given by the discontinuity of $H(h)$ along the support 
\ben
 \tilde u(h)=\displaystyle{\lim_{\gep\rightarrow 0}}\, \lB-{1\over 2\pi i}\left[H(h+i\gep)-H(h-i\gep)\right]\rB 
\een
for $h\in(q,r)$.
\end{enumerate}
We now employ the strategy prescribed in \cite{duttadutta} to find $H(h)$. In this branch Young distribution is non-zero between $q$ and $r$, therefore the resolvent should have a square root branch cut in that interval. We take the following ansatz for the resolvent
\begin{eqnarray}
H(h)=2\lnb{g(h)^2-\displaystyle{\sqrt{g(h)^2-f(h)^2}}\over v(h)},
\end{eqnarray}
with the constraint that the term inside the square-root can at most be a polynomial of degree $two$. Using property $(d)$ we have
\begin{eqnarray}
\displaystyle{\lim_{\gep\rightarrow 0}}\,\left[H(h+i\gep)+H(h+i\gep)\right]=2\lnb{f(h)^2\over v(h)^2}=2\lnb{h-p\over \bo}.
\end{eqnarray}
Therefore
\begin{equation}
f(h)=v(h)\, \sqrt{h-p\over \bo}.
\end{equation}
Using the above relation together with the asymptotic form of $H(h)$ we find
\begin{eqnarray}
g(h)=h+\bo -\half{1}-\half{p},\quad v(h)=2\bo,\quad f(h)= 2\sqrt{\bo(h-p)}.
\end{eqnarray}
Young distribution $\tilde u(h)$ is given by,
\be\label{eq:uhupcap}
\tilde u(h) = 1-{1\over \pi} \cos^{-1}\lB 1- {(h-q)(r-h)\over 2\bo(h-p)}\rB, \quad q\leq h\leq r
\ee
where,
\begin{eqnarray}
q&=&\bo -\frac{1}{2 \lambda }+\frac{3}{2}-\sqrt{{2\,\bo\over \gl}  (1-\lambda)},\\
r&=&\bo -\frac{1}{2 \lambda }+\frac{3}{2}+ \sqrt{{2\,\bo\over \gl}  (1-\lambda)}.
\end{eqnarray}
A typical Young distribution is plotted in figure \ref{fig:uhupcap}.
%%%%%%%%%%%%%%%%%%%%%%%%%%%%%%%%%%%%%%%%%%%%%%%%%%%%%
%%%%%%%%%%%%%%%%%%%%%%%%%%%%%%%%%%%%%%%%%%%%%%%%%%
%%%%%%%%%%%%%%%%%%%%%%%%%%%%%%%%%%%%%%%%%%%%%%%%%%%%%
%%%%%%%%%%%%%%%%%%%%%%%%%%%%%%%%%%%%%%%%%%%%%%%%%%

Here we see that the distribution has a spread $r-q = 2\sqrt{{2\,\bo\over \gl}  (1-\lambda)}$. Demanding that the spread is always less than $1\over \l$ we obtain
\be
\bo\leq {1\over 8\l(1-\l)},
\ee
which gives the upper boundary (figure \ref{fig:fullphase}) of the region of validity for this phase in $(\bo,\l)$ parameter space. Also, this phase is different than no-gap phase if $q\ge p$ which implies
\be
\bo\geq {1\over 2\l}-\half1.
\ee
This relation gives the other boundary in the parameter space.

\subsection{Duality between integrable representations}
\label{sec:dualityintreps}

The Young diagram for upper-cap phase has non-empty\footnote{For this phase $h_i =n_i +k-N -i$, since this is a representation for $SU(k-N)_N$.} rows from $y=0$ to $y=1/\l-1-p$. Since $0<p<1$ (see equation \ref{eq:pvalueuppcap}), this verifies that number of rows in this representation is always less than $k-N$. Similarly we calculate number of boxes in the first row for this diagram and it turns out to be $n_1 = N\lb r- \lb 1/\l -1\rb\rb$. It is easy to show that  if $\bo>1/2\l -1/2$ then $n_1<N$. This confirms that this representation corresponds to an integrable representation of $SU(k-N)_k$. Thus we explicitly check that level-rank duality maps integrable representations of $SU(N)_k$ to integrable representations of $SU(k-N)_k$.

\subsection{Phase space description } 

\begin{figure}[h]
	\centering
	\begin{subfigure}{0.4\textwidth}
		\centering
		\includegraphics[width=6cm,height=8cm]{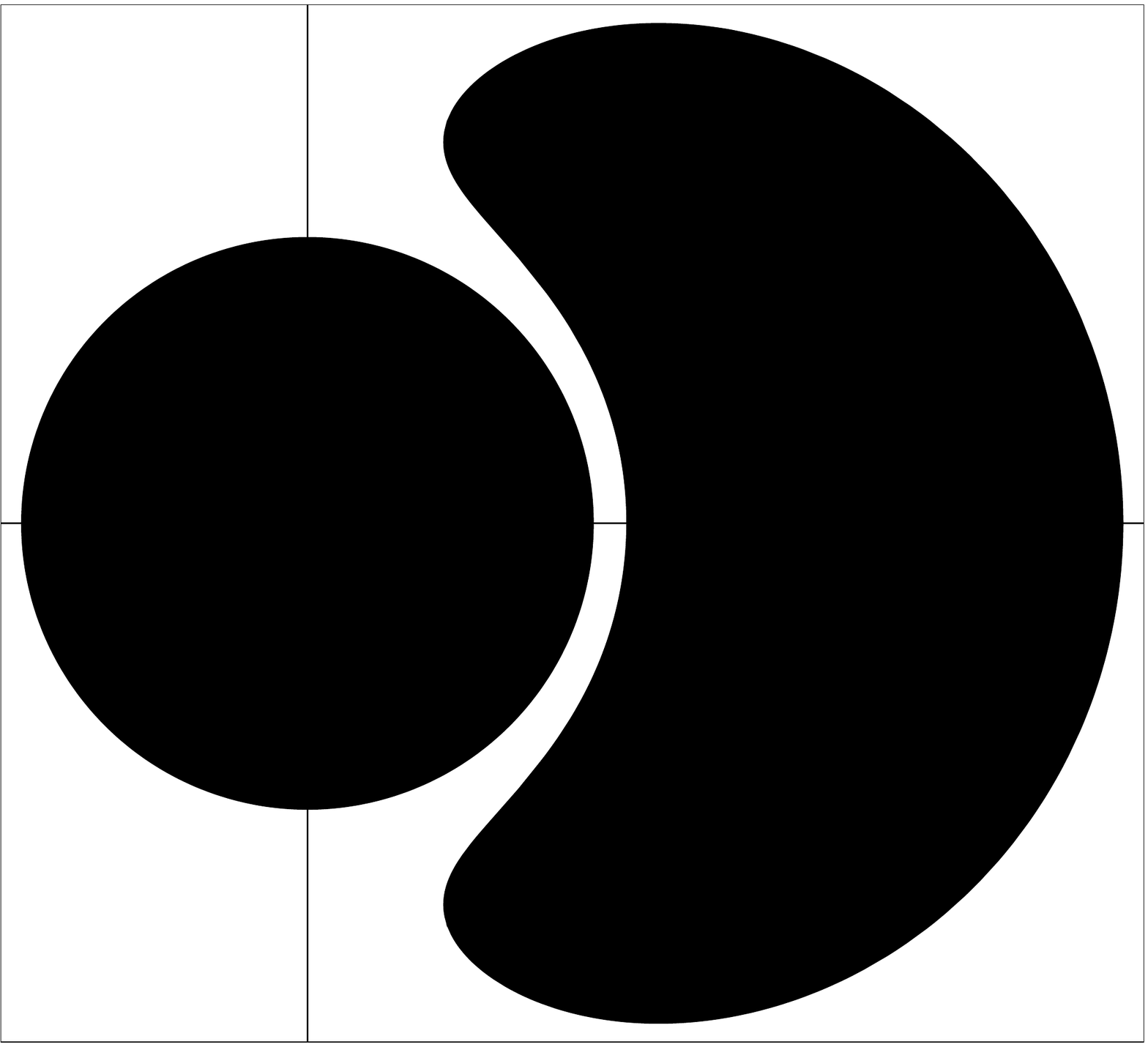}
		%		\caption{Phase space distribution for no-gap phase.}
	\end{subfigure}%
	\caption{Phase space distribution for upper-cap solution.}
	\label{fig:uppercap}
\end{figure}

As discussed in a series of papers \cite{duttagopakumar,duttadutta,riemannzero,Chattopadhyay} that momentum (Young diagram) representation and eigenvalue representation of large $N$ phases of any unitary matrix model can be unified in phase space representations (see section \ref{sec:youngreview}). To provide a phase space description for this phase,  we first need to understand how Young distribution and eigenvalue distribution are related to each other. We  invert the equation (\ref{eq:uhupcap}) and find, like one-gap phase, the functional inverse is not one-to-one. For a given $u(h)$ there exists two possible values of $h$,
\begin{eqnarray}
h_{\pm}=\half{1}+\bo \cos \gt+\half{p}\pm 2\, \bo  \sqrt{{{1\over\gl}-1\over 2\bo}-\sin^2{\pi u(h)\over 2}}\cos {\pi u(h)\over 2}.
\end{eqnarray}
Identifying $\pi \tilde u(h) = \q$ we define a phase space density
\be
\label{eq:phspdistriuppergap}
\o(h,\q) = {1\over 2\pi}\Theta\lb2-{1\over \l} -h\rb + {1\over 2\pi} \Theta(h-h_-(\q))\Theta(h_+(\q)-h).
\ee
Following (\ref{eq:rhoufromps}), we find that the above phase space distribution gives Young distribution (\ref{eq:uhupcap}). Eigenvalue distribution obtained from this distribution is given by,
\be
\bar \r(\q) = {h_+(\q)-h_-(\q)\over 2\pi}={1\over 2\pi}\lb 2- {1\over \l}\rb + {\bo\over 2\pi} \sqrt{{{1\over\gl}-1\over 2\bo}-\sin^2{\q\over 2}} \cos {\q\over 2}.
\ee
This eigenvalue distribution is related to eigenvalue distribution for upper-cap phase by an overall $\l$ dependent shift and a $\pi$ shift in $\q$
\be
\r(\q) = {1\over \pi}\lb {1\over \l -1} \rb + \bar\r(\q+\pi).
\ee
In figure \ref{fig:uppercap} we plot phase space distribution function \ref{eq:phspdistriuppergap} for upper-cap phase. 
%%%%%%%%%%%%%%%%%%%%%%%%%%%%%%%%%%%%%%%%%%%%%%%%%%%%%
%%%%%%%%%%%%%%%%%%%%%%%%%%%%%%%%%%%%%%%%%%%%%%%%%%
%%%%%%%%%%%%%%%%%%%%%%%%%%%%%%%%%%%%%%%%%%%%%%%%%%%%%
%%%%%%%%%%%%%%%%%%%%%%%%%%%%%%%%%%%%%%%%%%%%%%%%%%
The droplet corresponding to upper-cap is topologically different than droplets of no-gap and lower-gap phase (figure \ref{fig:phasespacenoonegap}). Here we see that phase space distribution has two disconnected droplets. This implies that Young diagram has a $\cO(N)$ jump in number of boxes.

\subsection{Level-rank duality and transposition of representations}
\label{sec:levelrankduality}

The level-rank duality in terms of renormalised level $k$ and rank $N$ is given by $N\ra k-N$ and $k\ra k$. As a result, under level-rank duality, the 't Hooft coupling constant $\l$ transforms as
\be
\label{eq:paramdual1}
\gl^D={k-N\over k}=1-\gl, \quad \text{$\gl^D$ is the 't Hooft coupling in dual theory.}
\ee
Demanding partition function is invariant under level-rank duality we find the second coupling constant $\bo$ also transforms under level-rank duality as
\be
\label{eq:paramdual2}
\gb_1^D={\gl\over 1-\gl}\gb_1, \quad \text{$\gb_1^D$ is the coupling in dual theory.}
\ee

It was shown in \cite{shirazs2s1} that under level rank duality the eigenvalue densities for lower-gap and upper-cap phase are related to each other by,
\begin{eqnarray}\label{eq:rhodual}
\rho_{uc}\left(\gl^D,\bo^D,\gt\right)={\gl\over 1-\gl}\left[{1\over 2\pi\gl}-\rho_{lg}(\gl,\bo,\gt+\pi)\right].
\end{eqnarray}
Therefore we expect that Young distributions for lower-gap phase (\ref{eq:uhonegap}) and upper-cap phase (\ref{eq:uhupcap}) are also related to each other by level-rank duality. To check this explicitly, we note that the continuous parameters $x$ defined in large $N$ limit (equation \ref{eq:contvardef}) for two different gauge groups $SU(N)$ and $SU(k-N)$ are given by
\be
x={i\over N},\quad \text{for $SU(N)$}\quad \text{and} \quad y={i\over k-N} \quad \text{for $SU(k-N)$}
\ee
and they are related to each other by
\begin{eqnarray}
y&=&{i\over k-N}={\gl\over 1-\gl}x.
\end{eqnarray}
Since the total number of boxes will remain same under transposition, we have
\begin{eqnarray}
N\int_0^1 n(x)\, dx&=&(k-N)\int_0^1 n^D(y)\,dy
\end{eqnarray}
which implies,
\begin{eqnarray}
n^D(y)={\gl\over 1-\gl} n(x).
\end{eqnarray}
Therefore the hook length in dual theory $h^D$ is related to hook length in original theory $h$ by
\begin{eqnarray}
\label{eq:hdual}
\begin{split}
h^D&=n^D+1-y \ = \ {\gl\over 1-\gl} n+1-{\gl\over 1-\gl}x\\
%&={\gl\over 1-\gl} (n(x)+1-x+p)\quad\text{where } p=2-{1\over \gl}\\
&={\gl\over 1-\gl}(h-p),\quad p=2-{1\over \gl}.
\end{split}
\end{eqnarray}
Using (\ref{eq:paramdual1},\ref{eq:paramdual2}) and (\ref{eq:hdual}) one can show that Young distribution for lower-gap (\ref{eq:uhonegap}) can be exactly mapped to Young distribution for upper-cap (\ref{eq:uhupcap}).

\section{Discussion and outlook}
\label{sec:discussion}

In this paper we explicitly find out large $N$ representations corresponding to no-gap, lower-gap and upper-cap phases of $SU(N)$ Chern-Simons gauge theory with level $k$ coupled with fundamental matter. We consider Gross-Witten-Wadia potential for fundamental matter. We find that representations corresponding to no-gap and lower-gap phase are integrable representations of $SU(N)_k$, whereas for upper-cap phase, representation are the integrable representations of $SU(k-N)_k$. We also prove that, similar to eigenvalue density function, Young diagram distribution functions corresponding to lower-gap and upper-cap phases are also related to each other by level-rank duality. Therefore, our result verifies that level-rank duality is actually a map between integrable representations of two dual theories. The constraints on representations follows from the fact that eigenvalue densities for these phases are restricted by an upper cap. Our observation is based on the result of \cite{duttagopakumar} which found an identification between eigenvalue density and Young diagram distribution functions for different phases in the large $N$ limit.

Large $N$ representation for lower-gap with upper-cap phase is still missing in our work. Since GWW potential is self dual, lower-gap with upper-cap phase is also dual to itself. Therefore, knowledge of level-rank duality will not be helpful to construct this representation. We need to solve saddle point equation to find the dominant representation from the first principle. Work of \cite{duttadutta} will give helpful hint to find a solution. Since eigenvalue density has two gaps for this phase, we guess that Young diagram distribution will also have two gaps, which implies the resolvent defined by equation (\ref{eq:resolventH}) will have two branch cuts. However, analytically computing such resolvent seems to be very cumbersome and also it becomes difficult to find identification between Young diagram side and eigenvalue side. We are working on this problem.

In this paper we have considered GWW potential. One can also extend phase space formalism to find out large $N$ representations for other fundamental matter Chern-Simons theories which are not necessarily self-dual. In the high temperature limit the duality between the CS theory coupled with fermions and bosons have already been proved by \cite{takimi2013,shirazs2s1}. In this case the lower gap phase of one theory maps to the upper cap phase of the dual theory under level-rank interchange. The potential for fundamental matter can be written a generic single plaquette model. Corresponding partition function can also be written as sum over representations of unitary group. But computation of character of symmetric group turns out to be very difficult in this case, as it involves all possible cycles. There exists no exact formula for character of symmetric group in terms of number of cycles and number of boxes for a given representation. One can use Frobenius formula to write character in terms of auxiliary variables \cite{Chattopadhyay}. We are currently looking at this issue.

The phase space description is also interesting for the upper cap phase. As already pointed out by \cite{duttagopakumar}, the different large $N$ phase of any theory can be described in terms of different topologies of droplet. It has already been established that for no-gap phase the origin is always inside the droplet and for the one-gap/lower gap phase origin can at most\footnote{Reducing the Young diagram amounts to moving the droplet towards the origin. The reduced diagram corresponds to a droplet with origin on its left boundary.} be the limit point of the corresponding droplet. These topologies turned out to be very robust \cite{Chattopadhyay}. In this paper we found that the droplet picture for the upper cap phase forms two disconnected islands, one includes the origin. Though currently we don't explicitly have the actual droplet picture for the upper cap with a lower gap phase, but depending on our formalism one may speculate the droplet picture for the two gap phase as well.

\bc
---------------------------
\ec

\paragraph{Acknowledgement} We would like to thank Rajesh Gopakumar for many helpful discussion. We are grateful to D. Ghoshal, S. Jain, S. Minwalla, S. Mukhi for discussion. AC would like to acknowledge the hospitality of IISER Pune and ICTS Bangalore where part of this work is done. SD acknowledges the Simons Associateship, ICTP. Work of SD is supported by DST under a project with number {\it EMR/2016/006294}. Finally, we are indebted to people of India for their unconditional support towards researches in basic sciences. 

\appendix

\section{Young Diagrams and Integral representations of affine Lie algebra}\label{app:yng}
\subsection{Young Diagrams}
We will skip the preliminary discussions on Young diagrams for unitary groups\footnote{\cite{sojagroup} is a good reference for interested readers.}. For a given affine Lie algebra of the form $A_n$, the extended Dynkin diagrams have two different symmetries. One is the $\mathcal{Z}_2$ reflection symmetry and the second one is the $\mathcal{Z}_N$ cyclic symmetry. This two symmetries are responsible for the conjugate diagrams and the definition of cominimal equivalence class. Before mathematically establishing them it is best to pictorially understand some concepts first.

\paragraph{Transposition} As the name suggests transposition involves exchange of rows and columns of a given Young diagram. For example under the operation of transposition
\begin{figure}[H]
	\centering
	\includegraphics[width=9cm,height=5cm]{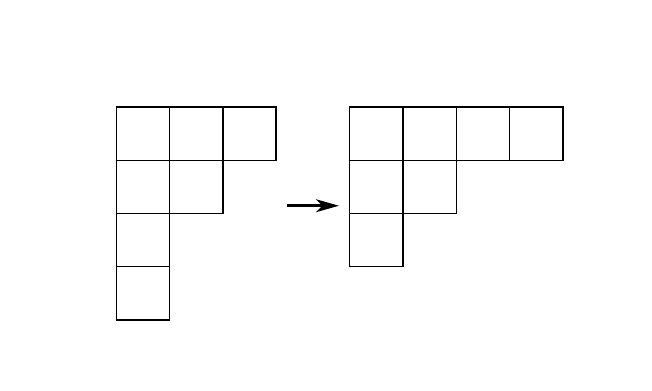}
\end{figure}

\paragraph{Conjugate Representation}The conjugate Young diagram corresponding to some given Young diagram can be obtained by the following simple way. For a $SU(N)$ group replace all $i$ number of boxes in a column by $(N-i)$ number of boxes and flip the resulting diagram vertically. Let us take the case of $SU(4)$ for illustration\\
\begin{figure}[H]
	\centering
	\includegraphics[width=\textwidth]{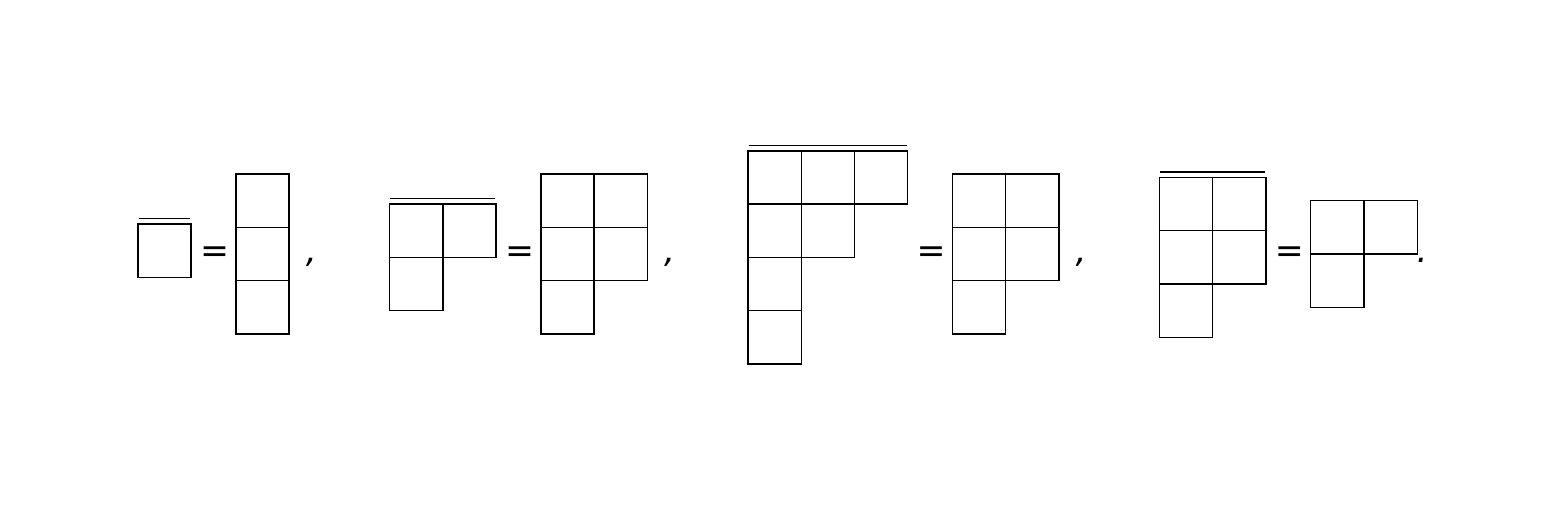}
\end{figure}
%
%
%\begin{eqnarray}
%\begin{matrix}\overline{ \ydiagram[*(white) ]{1}}\end{matrix}
%= \begin{matrix}\ydiagram[*(white)]{1,1,1}\end{matrix}\,\,\, ,\qquad
%\bnm \overline{ \ydiagram[*(white)]{2,1} }\enm=\bnm \ydiagram[*(white)]{2,2,1}\enm\,\,\, ,\qquad \bnm %\overline{ \ydiagram[*(white)]{3,2,1,1} }\enm=\bnm \ydiagram[*(white)]{2,2,1}\enm\,\,\, , \qquad \bnm %\overline{ \ydiagram[*(white)]{2,2,1}}\enm=  \bnm \ydiagram[*(white)]{2,1}  \enm.
%\end{eqnarray}
%
From the second and third diagram it is clear that representations which only differ by columns of length $N$ attached to the left have equivalent conjugate representation. Now looking at the last two diagrams one can convincingly understand that the diagrams differing by columns of length $N$ attached to the left are equivalent representation. Therefore any Young diagram for $SU(N)$ can be expressed in the ``reduced" form, where the bottom row will always be empty.

\subsection{Integrable representations}
\begin{figure}[H]
	\centering
	\includegraphics[width=0.8\textwidth]{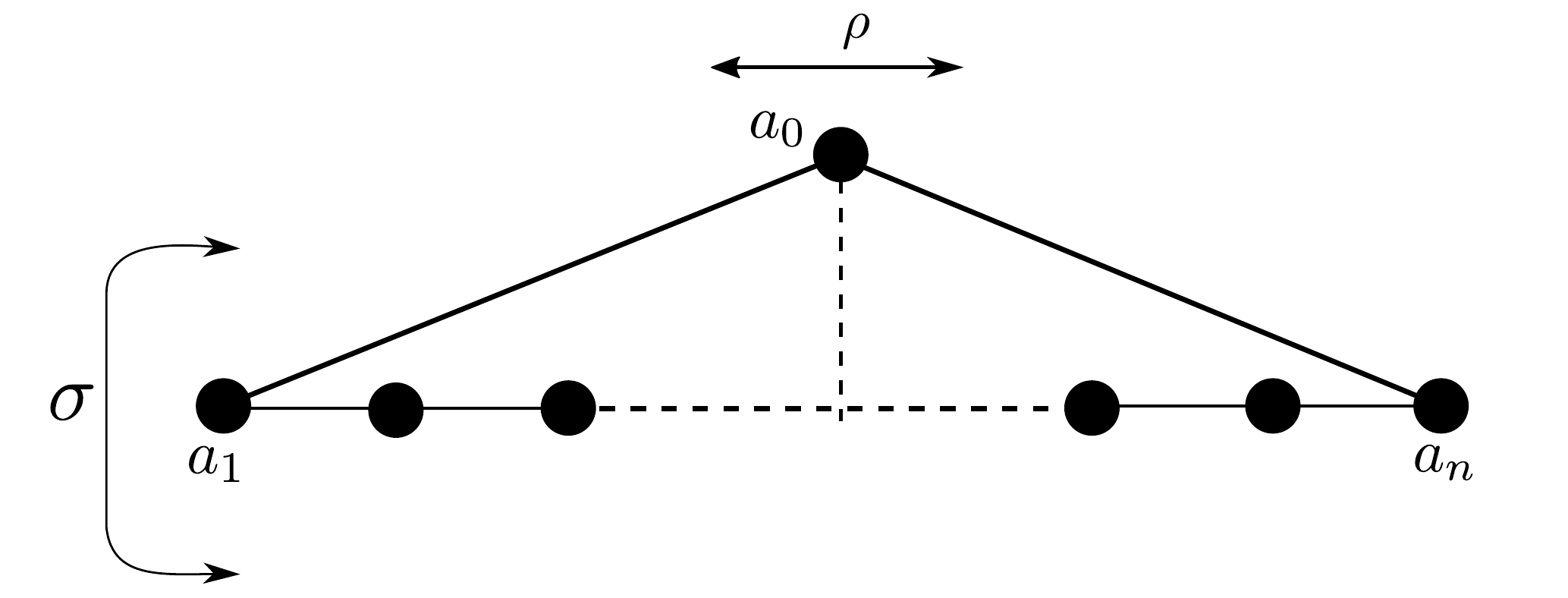}
	\caption{extended Dynkin diagram}
	\label{fig:dynkinkin}
\end{figure}

The affine extension of any Lie algebra can be realised by adding an extra node in the corresponding Dynkin diagram related to the highest root of the previous one. This extra node has the affect of making the root system infinite. This infinite dimensional highest weight representations can be organised in terms of \emph{"level"} of the algebra, which corresponds to the central extension of the corresponding loop algebra. A Lie algebra with rank $r$ has $r$ number of nodes in its Dynkin diagram, naturally the extended Dynkin diagrams has $r+1$ nodes. The \emph{level} of the affine Lie algebra is defined as the sum of all its Dynkin labels multiplied by its corresponding comark. More on this related topics can be found in \cite{yellowbook}. Here we will only demonstrate the properties and symmetries of the \footnote{here the convention will be $SU(rank)_{level}$ for convenience. The affine algebra will be denoted by an upper hat.} $\widehat{su}(N)_k$ affine Lie algebra. The extended Dynkin diagram is shown in figure \ref{fig:dynkinkin}.
Here $a_0$ is the extra Dynkin label defined as 
\begin{eqnarray}
a_0=k-\sum_{i=1}^r m_ia_i\,\,\,; \qquad m_i\Longrightarrow\text{Components of the highest co-root.}
\end{eqnarray}
Representations that can be decomposed into finite irreducible copies of $su(2)$ are called  integrable representations. An integrable highest weight representation has $a_0\geq 0$. The row lengths of a \emph{reduced} Young diagram corresponding to an irreducible representation of $SU(N)$ can be written as
\begin{eqnarray}
l_i=\Bigg\{ \begin{matrix}
	\sum_{j=i}^{N-1}a_j\qquad j\in(1,N-1) \\ 0\qquad i=N
	\end{matrix}
\end{eqnarray}
The condition of integrable representations now restricts that the reduced Young diagram must have $l_1\leq k$. This extended Dynkin diagrams has $\mathbb{Z}_2$ as well as $\mathbb{Z}_N$ symmetry. The $\mathbb{Z}_2$ symmetry $\rho$ takes a representation $y$ to its conjugate representation $\rho(y)=\bar{y}$, where
\begin{eqnarray}
\bar{y}_i=
\Bigg\{ \begin{matrix}
y_{N-i}\qquad i\in(1,N-1) \\y_0\qquad i=0
\end{matrix}.
\end{eqnarray}
The additional $\mathbb{Z}_N$ cyclic symmetry $\sigma$ of the extended Dynkin diagrams takes a representation $y$ into $\tilde{y}=\sigma(y)$, where 
\begin{eqnarray}
\tilde{y}_i=
\Bigg\{ \begin{matrix}
y_{i-1}\qquad i\in(1,N-1) \\y_{N-1}\qquad i=0
\end{matrix}.
\end{eqnarray}
As one can see that this operation basically adds one extra row of width $k$ to the top of the Young diagram (reduced), therefore the transpose of both $y$ and $\tilde{y}$ corresponds to the same representation in the dual $SU(k)_N$ theory by the reduction rule for $SU(k)$ Young diagrams. Thus $y$ and $\tilde{y}$ are called "\emph{cominimally equivalent}".

\bibliography{bibfors2s1}{}
\bibliographystyle{jhep}

\end{document}